\documentclass[prd,aps,a4paper,eqsecnum,nofootinbib,floatfix,twocolumn]{revtex4}  %

\usepackage{graphicx} 
\usepackage{amsmath,amsfonts,amssymb}
\newcommand{\beq}{\begin{equation}}
\newcommand{\eeq}{\end{equation}}
\newcommand{\bea}{\begin{eqnarray}}
\newcommand{\eea}{\end{eqnarray}}
\newcommand{\be}{\begin{equation}}      
\newcommand{\ee}{\end{equation}}

\def\nn{\nonumber}

\begin{document}

\title{Scalar perturbations in a Top-Star spacetime}

\author{Massimo Bianchi$^{1}$, Donato Bini$^{2,3}$, Giorgio di Russo$^{1,4}$}
  \affiliation{
$^1$Dipartimento di Fisica, Universit\`a di Roma \lq\lq Tor Vergata" and Sezione INFN Roma2, Via della
Ricerca Scientifica 1, 00133, Roma, Italy\\
$^2$Istituto per le Applicazioni del Calcolo ``M. Picone,'' CNR, I-00185 Rome, Italy\\
$^3$INFN, Sezione di Roma Tre, I-00146 Rome, Italy \\
$^4$Institut de Physique Th\'eorique, Universit\'e Paris Saclay, CEA, CNRS, 
F-91191 Gif-sur-Yvette, France}

\date{\today}

\begin{abstract}
We discuss the dynamics of a (neutral) test particle in Topological Star spacetime undergoing scattering processes by a superposed test radiation field, a situation that in a 4D black hole spacetime is known as relativistic Poynting-Robertson effect, paving the way for future studies involving radiation-reaction effects.
Furthermore, we study self-force-driven evolution of a scalar field,   perturbing the Top-Star spacetime with a scalar charge current. The latter for simplicity is taken to be circular, equatorial and geodetic. To perform this study, besides solving all the self-force related problem (regularization of all divergences due to the self-field, mode sum regularization, etc.), we had to adapt the 4D Mano-Suzuki-Takasugi formalism to the present 5D situation.
Finally, we have compared this formalism with the (quantum) Seiberg-Witten formalism, both related to the solutions of a Heun Confluent Equation, but appearing in different contexts in the literature, black hole perturbation theory the first, quantum curves in super-Yang-Mills theories the second.
\end{abstract}

\maketitle

\section{Introduction}

Discriminating Exotic Compact Objects (ECOs) from General Relativity (GR) Black Holes (BHs) is an active endeavour \cite{Bianchi:2020miz,Bianchi:2020bxa,Ikeda:2021uvc,Bianchi:2021xpr,Bianchi:2021mft,Consoli:2022eey,Bah:2021jno,Bena:2020uup,Bena:2020see,Heidmann:2022ehn,Fucito:2023afe,DiRusso:2024hmd}. One possible test to perform is to analyze the Gravitational Wave (GW) signal emitted by a low-mass probe slowly inspiralling into a very massive central body, that is often called an Extreme Mass Ratio Inspiral (EMRI). For the same mass, angular momentum and charge, the GW signal emitted by ECOs  is expected to produce tails of echoes in the ring-down phase due to long-lived  quasi normal modes (QNMs) associated with the presence of internal `meta-stable' photon-rings and to the absence of a horizon \cite{Barack:2018yly,Berti:2015itd,Abedi:2016hgu,Berti:2009kk}. Actually, even the prompt ring-down QNMs associated with the unstable photon-rings of ECOs may slightly differ from the QNMs of BHs with the same asymptotics.  It is therefore legitimate to ask whether GWs emitted during the inspiral phase can teach us something about  ECOs and allow us to discriminate them from putative BHs.

A particularly simple class of ECOs are  `topological stars' (top stars or TS in short), discovered by Bah and Heidemann  \cite{Heidmann:2022ehn,Bah:2022yji,Bah:2020pdz,Bah:2020ogh,Bah:2021irr}  in the framework of the  fuzzball proposal \cite{Lunin:2001jy,Mathur:2005zp,Bena:2007kg,Skenderis:2008qn}. In this approach, the objects colloquially called BHs are to be regarded as ensembles of smooth horizonless geometries \cite{Lunin:2001jy,Mathur:2005zp,Bena:2007kg,Skenderis:2008qn}, with the same asymptotics as their relatives in GR\footnote{An alternative emerging approach employs matter shells in the BH interior to describe BH microstates with horizons and singularity at least at the semi-classical level \cite{Balasubramanian:2022lnw,Balasubramanian:2022gmo,Climent:2024trz}.}. It is well known that this is only possible in $D>4$; in particular, Top(ological) Stars (TS)
are smooth and horizonless static solutions of Einstein-Maxwell theory in $D=5$ that reduce to spherically symmetric solutions of Einstein-Maxwell-Dilaton theory in $D=4$.  They have received special attention in the last couple of years  \cite{Heidmann:2022ehn,Bianchi:2023sfs,Heidmann:2023ojf,Cipriani:2024ygw,Bena:2024hoh,Dima:2024cok}  for their `resemblance' to spherically symmetric (charged) BHs coupled to a scalar `dilaton'.

Aim of the present investigation is to elaborate on these recent works, and determine the waveform of scalar waves -- as a proxy of GWs -- emitted by a low-mass ($\mu \ll M_{\rm TS}$) neutral scalar (non-spinning) particle probing a TS. To be specific, the probe will be taken to be on a circular orbit at $r_0\gg G M_{\rm TS}$. To this end we have to tackle the linearized wave equation for the emitted radiation and identify the relevant geodesics for the probe.

Thanks to spherical symmetry, separation of the angular dynamics in both cases leads to integrable radial equations. For the low-mass geodesics, imposing circular orbits makes the analysis almost straightforward. While for the (massless) waves one has to face with a Confluent Heun Equation (CHE) as for all (asymptotically flat) BHs in $D=4$. This can be solved in two different but equivalent ways. 

The first one is the approach pioneered by Leaver \cite{leaver1,leaver2}, and then further consolidated by Mano, Suzuki, Takasugi (MST), Sasaki, Tanaka \cite{Mano:1996mf,Mano:1996vt,Sasaki:2003xr,Mino:1997bx}
that amounts to introduce a `renormalized' angular momentum $\nu = \ell + ...$ and express the solution, a Confluent Heun Function (CHF), as a series of hypergeometric functions around the `horizon' or the `cap' (a regular singular point) and as an infinite series of confluent hypergeometric functions (or Coulomb functions, to be more precise) around infinity (an irregular singular point). In the resulting Post-Minkowskian (PM) approximation the `small' (and dimensionless) expansion parameter is $\varepsilon \approx \omega M$ with $\omega$ the frequency of the wave and $M$ the mass of the BH or ECO.

The second approach \cite{Aminov:2020yma,Aminov:2023jve} 
is based on the observation that the relevant CHE can be regarded as the quantum Seiberg-Witten \cite{Seiberg:1994aj,Seiberg:1994rs}
curve of an $N=2$ super Yang-Mills (SYM) theory, with gauge group $SU(2)$ coupled to $N_f=3$ hypermultiplets in the fundamental `doublet' representation of the gauge group in a Nekrasov-Shatashvili (NS) \cite{Nekrasov:2009rc,Nekrasov:2002qd} non-commutative background.
In this framework, the PM parameter  $\varepsilon$ is (related to) the instanton counting parameter $q$, which in turn is related to the `gauge coupling' or better to the RG invariant scale by $q=\Lambda^\beta$ with $\beta= 2N_c-N_f =2\times 2 -3 =1 $ in the cases of present interest\footnote{In \cite{Bianchi:2021xpr,Bianchi:2021mft,Consoli:2022eey,Bianchi:2023rlt,Bianchi:2022qph,DiRusso:2024hmd} 
we have extended the qSW-BH/ECO correspondence to many other cases with $4\ge N_f\ge 0$.}. Quite remarkably the qSW  description identifies the `renormalized' angular momentum $\nu$ with the quantum $a$-cycle $a(q, u; m_f)$; more precisely, 
\be
\nu = a -{1\over 2}\,.
\ee
We will give indirect proof or rather offer evidence of this identity later on, although most of our results are obtained within the first approach. Notice that for scalar perturbations $\nu$ is an even function of $\varepsilon$. This is straightforward to see in MST approach, while it is not at all obvious to see in the qSW approach, even though the relevant recursion relations look less laborious. 

The plan of the paper is as follows.

In Section II, after a brief review of TS and their basic properties, we revisit the geodesics and identify the low-mass `critical' geodesics (constant $r$).
In Section III, we study non-geodetic motion due to a superposed test field, both in the cases of a radiation field and of test fluid (dust, for simplicity).
In Section IV and V, we write down the (massless, neutral) wave equation which we analyze in the source-free situation (Sec. IV) for which we explicitly provide various types of \lq\lq In" and \lq\lq Up" solutions (Sec. V), namely simple Post-Newtonian (PN) solutions, Mano-Suzuki-Takasugi (MST)-type and JWKB-type.
The latter types of solutions are then used in Section VI and VII to discuss the scalar self-force problem on a TS background, {\it i.e.} computing back-reaction effects due to a scalar charge moving along a circular orbit.
Finally, we relegate to Appendix A the presentation of the quantum Seiberg-Witten (qSW) approach to the radial part  of wave equation, which turns out to be a CHE, showing agreement between the MST types of solutions and the qSW one up to a certain PN order (up to which we could compute both types of solutions), modulo an overall normalization factor.
Whenever it is possible we add a comparison with well-known results for `standard' BHs. In fact, for $r_b=0$ (and thus $r_s=2M$ and $Q_S=0$) the results coincide with the ones for a Schwarzschild BH. We will often rely on this limit as a sanity check in the following. 

\section{Top star spacetime metric}
\label{TSmetric}

A Topological Star is described by the following metric in $D=5$\footnote{We use mostly positive signature, $-++++$, and coordinates $x^\mu=\{t,r,\theta,\phi,y\}$, $\mu=0,\ldots 4$.}
\bea
\label{metric_top_star}
ds^2 &=& -f_s(r)dt^2+\frac{dr^2}{f_s(r)f_b(r)} \nonumber\\ 
&+& r^2 (d\theta^2+\sin^2\theta d\phi^2) + f_b(r) dy^2\,,
\eea
with 
\beq
f_{s,b}(r)=1-{r_{s,b}\over r}\,.
\eeq
The coordinate $y$ is compact $y \sim y+ 2\pi R_y$. 
The geometry \eqref{metric_top_star} represents a magnetically charged solution of 5D Einstein-Maxwell's equations, sourced by the electromagnetic field
\beq
F=P\sin\theta d\theta \wedge d\phi\,,\qquad P=\frac{3r_br_s}{2\kappa_5^2}\,, 
\eeq
with $P$ representing a `magnetic' charge\footnote{An `electric' solution corresponding to a string wound around the $y$ direction with $H_3 = Q dt\wedge dy\wedge dr/r^2={}^{*_5} F_2$ is also known.}.
From a 5D perspective the geometry is regular for $r_s$, $r_b$ and $R_y$ satisfying
the condition
\beq
r_s=r_b \left(1-\frac{4r_b^2}{R_y^2}\right)
\eeq
that implies that $r_s \le  r_b$ and $R_y \ge  2r_b$\footnote{At fixed mass $M_{\rm TS}$ the minimal radius is $R_y = 4\sqrt{2}M_{\rm TS}$ reached for $r_b=2r_s$.}.
Solutions without thermodynamical (or Gregory-Laflamme) instabilities require $r_s < r_b < 2r_s$ and in general belong to two different classes (see e.g. \cite{Bianchi:2023sfs}).
After dimensional reduction to $D=4$, the solution exposes a naked singularity and has a mass
\beq
\label{mass_TS}
4 G_4 M_{\rm TS} = 2 r_s + r_b\,,
\eeq
with 
\beq
8\pi G_4 =\kappa_4^2= {\kappa_5^2\over 2\pi R_y}=\frac{8\pi G_5}{2\pi R_y}\,. 
\eeq
Hereafter we will   set $G_4=1=c$.
For $r_b=0$, and thus $r_s=2GM_{\rm TS}$, the resulting singular solution is a Schwarzschild BH times a circle.
We will often resort to this limit in order to check the results. 
The Kretschmann invariant for this spacetime reads
\bea
\label{krets_TS}
{\mathcal K}&=&R^{\mu\nu\rho\sigma}R_{\mu\nu\rho\sigma}\nonumber\\
&=&\frac{12 (r_s r_b+r_b^2+r_s^2)}{r^6}-\frac{30 r_s r_b (r_s+r_b)}{r^7}\nonumber\\
&+&\frac{115 r_s^2 r_b^2}{4 r^8}\,,
\eea
and $r_b\not= 0$ is responsible for curvature contributions $O\left( \frac{1}{r^7}\right)$ and $O\left(\frac{1}{r^8}\right)$ absent in the 4D Schwarzschild case.
Note that, in general, for values of $r_b>r_s$,  the Kretschmann invariant ${\mathcal K}$ is monotonically decreasing as $r\to \infty$.
Similarly, the electromagnetic counterpart $F_{\mu\nu}F^{\mu\nu}$ is given by
\beq
{\mathcal I}=F_{\mu\nu}F^{\mu\nu}=\frac{3 r_b r_s}{r^4\kappa_4^2} \,.
\eeq

It is worth to recall explicitly the Kretschmann invariant for the Schwarzschild  spacetime 
\beq
{\mathcal K}_{\rm schw}=\frac{12 r_s^2}{r^6}=\frac{48M^2}{r^6}\,.
\eeq
In order to make a more meaningful comparison with Eq. \eqref{krets_TS} one may think of replacing the Schwarzschild mass $M$ with the mass of a TS, Eq. \eqref{mass_TS}, namely
\beq
{\mathcal K}_{\rm schw} \bigg|_{M=M_{\rm TS}}=\frac{3}{r^6}\left( 2r_s+ r_b\right)^2=\frac{12 r_s^2+12 r_s r_b+3r_b^2}{r^6}\,.
\eeq
However, even this expression differs from the corresponding 5D quantity, Eq. \eqref{krets_TS}, 
\beq
{\mathcal K}-{\mathcal K}_{\rm schw} \bigg|_{M=M_{\rm TS}}=\frac{9 r_b^2}{r^6}-\frac{30 r_s r_b (r_s+r_b)}{r^7}+\frac{115 r_s^2r_b^2}{4r^8} \,,
\eeq
also by terms $O(1/r^6)$. Such $O(1/r^6)$ terms in the difference can be eliminated only if in ${\mathcal K}_{\rm schw}$ one replaces
$M\to M_*=c_1 r_s+c_2 r_b$ with 
$$M_*=\frac{\sqrt{r_b^2+r_br_s+r_s^2}}{2}\,,$$
which is however of poor geometrical/physical significance.

\subsection{Geodesics}

Let $P=P^\mu\partial_\mu$ be the five momentum vector tangent to a timelike geodesic of the metric \eqref{metric_top_star}, related to the 5-velocity as $P=\mu U$, with $\mu$ the mass of the particle\footnote{We denote the mass by $\mu$ to avoid confusion with the azimuthal number $m$ that will appear (and soon disappear, thanks to spherical symmetry) later on.} and $U$ unit, timelike, and future oriented.
We find
\bea 
\label{conjcan}
P_t&=&-E=-f_s(r)\dot{t}\,,\nonumber\\
P_r&=& \frac{\dot{r}}{f_s(r)f_b(r)}\,,\nonumber\\
P_{\theta}&=& r^2\dot{\theta}\,,\nonumber\\
P_\phi&=& J=r^2\sin^2\theta \dot{\phi}\,,\nonumber\\
P_y&=& {p} = f_b(r)\dot{y}\,,
\eea
with  $P_t=-E$, $P_\phi=J$ and $P_y={p}$ constants of motion. Using the Hamiltonian formalism one has
\be
{\mathcal H}=\frac12 g^{\mu\nu}P_\mu P_\nu\,,
\ee
with ${\mathcal H}=-{\mu^2/2}$
the mass shell condition. Using Eqs. \eqref{conjcan}  we find
\bea
2\mathcal{H}+\mu^2&=&P_r^2f_s(r)f_b(r)-\frac{E^2}{f_s(r)}+\frac{{p}^2}{f_b(r)}+\mu^2\nonumber\\
&+&{1\over r^2}\left(P_\theta^2+{J^2\over \sin^2\theta}\right)\,,
\eea
from which the (covariant) radial and the polar angular momenta follow by separation of variables
\bea
\label{prpq}
P_r^2&=& Q_r(r)\nonumber\\
&=& \frac{1}{(r-r_b)^2(r-r_s)^2} \left\{r^2[r(r-r_b)E^2-r(r-r_s){p}^2\right.\nonumber\\
&-&\left. (r-r_b)(r-r_s)\mu^2]-(r-r_b)(r-r_s)K^2\right\}\,,\nonumber\\
P_\theta^2&=& Q_\theta(\theta)= K^2-{J^2\over \sin^2\theta}\,,
\eea
where $K^2$ is the (positive) separation constant, representing the square of the orbital angular momentum of the probe. 

Usually one defines the effective potential for the radial motion as
\bea
\label{veffts}
\dot{r}^2&\equiv &Q_{\rm geo}=E^2-{p}^2-\mu^2-V_{\rm eff}(r)\,,\nonumber\\
V_{\rm eff}(r)&=& {r_b E^2\over r}-{r_s {p}^2\over r}-\left({r_s+r_b\over r}-{r_sr_b\over r^2}\right)\mu^2\nonumber\\
&+& {1\over r^2}\left(1-{r_s\over r}\right)\left(1-{r_b\over r}\right)K^2\,.
\eea
Thanks to spherical symmetry, we can simplify the discussion without losing generality by focusing on equatorial geodesics, with $\theta=\pi/2$ and thus $\dot\theta= P_\theta=0$ and $K^2=J^2$ (from the second of \eqref{prpq}). 

Critical geodesics, also referred to as photon-spheres, or more broadly, light-rings\footnote{We will use this terminology for massive probes, too.}, are uniquely determined by the condition for turning points, $\dot{r}=0$, along with the additional requirement of zero radial acceleration, $\ddot{r}=0$. While four-dimensional BHs typically exhibit a single unstable photon-sphere (a circular geodesic located at the maximum of the effective potential), compact geometries often feature an outer unstable photon-sphere and an inner stable photon-sphere \cite{Bianchi:2022qph, Bianchi:2023rlt, DiRusso:2024hmd}. Massive probes can also move along innermost stable circular orbits (ISCOs) outside the external photon-sphere as we are going to show. Let us rescale  $V_{\rm eff}$ by $M^2\mu^2$ in \eqref{veffts}  and introduce the  notation\footnote{Recall that the mass of the TS is $ M_{\rm TS}=\frac12 (r_s+r_b/2)$.} 
\beq
r=M \hat{r}\,,\qquad r_{b,s}=M \hat{r}_{b,s}\,,
\eeq 
together with
\be
\hat{K}={K\over \mu M}=\hat{\kappa} \hat r_s\,, \qquad \qquad \hat{E}=\frac{E}{\mu}\,\,.
\ee
Critical geodesics are then found by solving
\be
Q_{\rm geo}(\hat r_c,\hat E_c)=\partial_r Q_{\rm geo}(\hat r_c,\hat E_c)=0\,.
\ee
In the simpler case ${p}=0$ the critical values are the following
\bea
\hat r_{c,1}&=&\hat r_b\,, \nonumber\\ 
\hat E_{c,1}&=&{\sqrt{(\hat r_b-\hat r_s)(\hat{\kappa}^2 \hat r_s^2+\hat r_b^2)}\over \hat r_b^{3/2}}\,,\nonumber\\
\hat r_{c,2}&=&\hat{\kappa} \hat r_s (\hat{\kappa} - \sqrt{\hat{\kappa}^2 -3})\,, \nonumber\\
\hat E_{c,2}&=&\frac{2}{3\sqrt{6}}\sqrt{\hat{\kappa}^2+9+\frac{1}{\hat{\kappa}}(\hat{\kappa}^2-3)^{3/2}}\,,\nonumber\\
\hat r_{c,3}&=&\hat{\kappa} \hat r_s (\hat{\kappa}  + \sqrt{\hat{\kappa}^2 -3})\,, \nonumber\\
\hat E_{c,3}&=&\frac{2}{3\sqrt{6}}\sqrt{\hat{\kappa}^2+9-\frac{1}{\hat{\kappa}}(\hat{\kappa}^2-3)^{3/2}}\,.
\eea
The values of the second derivative are respectively:
\bea
\partial_{\hat r}^2Q_{\rm geo}\bigg|_{(\hat r_{c,1},\hat E_{c,1})}&=& \mu^2 \left[\frac{4\hat{\kappa}^2 \hat r_s^2}{\hat r_b^5}\left(\hat r_b-\frac32 \hat r_s\right)-2\frac{\hat r_s}{\hat r_b^2}  \right]\,,\nonumber\\ 
\partial_{\hat r}^2Q_{\rm geo}\bigg|_{(\hat r_{c,2},\hat E_{c,2})}&=& -\frac23  \mu^2 \frac{\sqrt{\hat{\kappa}^2-3}}{\hat{\kappa}^4 \hat r_s^3}A_+\nonumber\\
\partial_{\hat r}^2Q_{\rm geo}\bigg|_{(\hat r_{c,3},\hat E_{c,3})}&=& -\frac23  \mu^2 \frac{\sqrt{\hat{\kappa}^2-3}}{\hat{\kappa}^4 \hat r_s^3}A_-\,,
\eea
where
\beq
A_\pm=\frac{ \hat r_b\sqrt{\hat{\kappa}^2-3}\pm\hat{\kappa} (\hat r_b-3\hat r_s) }{[\hat{\kappa}-\sqrt{\hat{\kappa}^2-3}]^4}\,,
\eeq
and derivatives are taken with respect to the rescaled radial variable $\hat r$.
For $r_b<\frac32 r_s$, $r_{c,1}$ and $r_{c,2}$ are a stable and unstable photon-sphere respectively, while  $r_{c,3}$ is always stable (ISCO). In the following we will mostly focus on the latter. Before doing this let us briefly comment on non critical geodesics.

\begin{figure*}[h]
\[
\begin{array}{cc}
\includegraphics[scale=0.30]{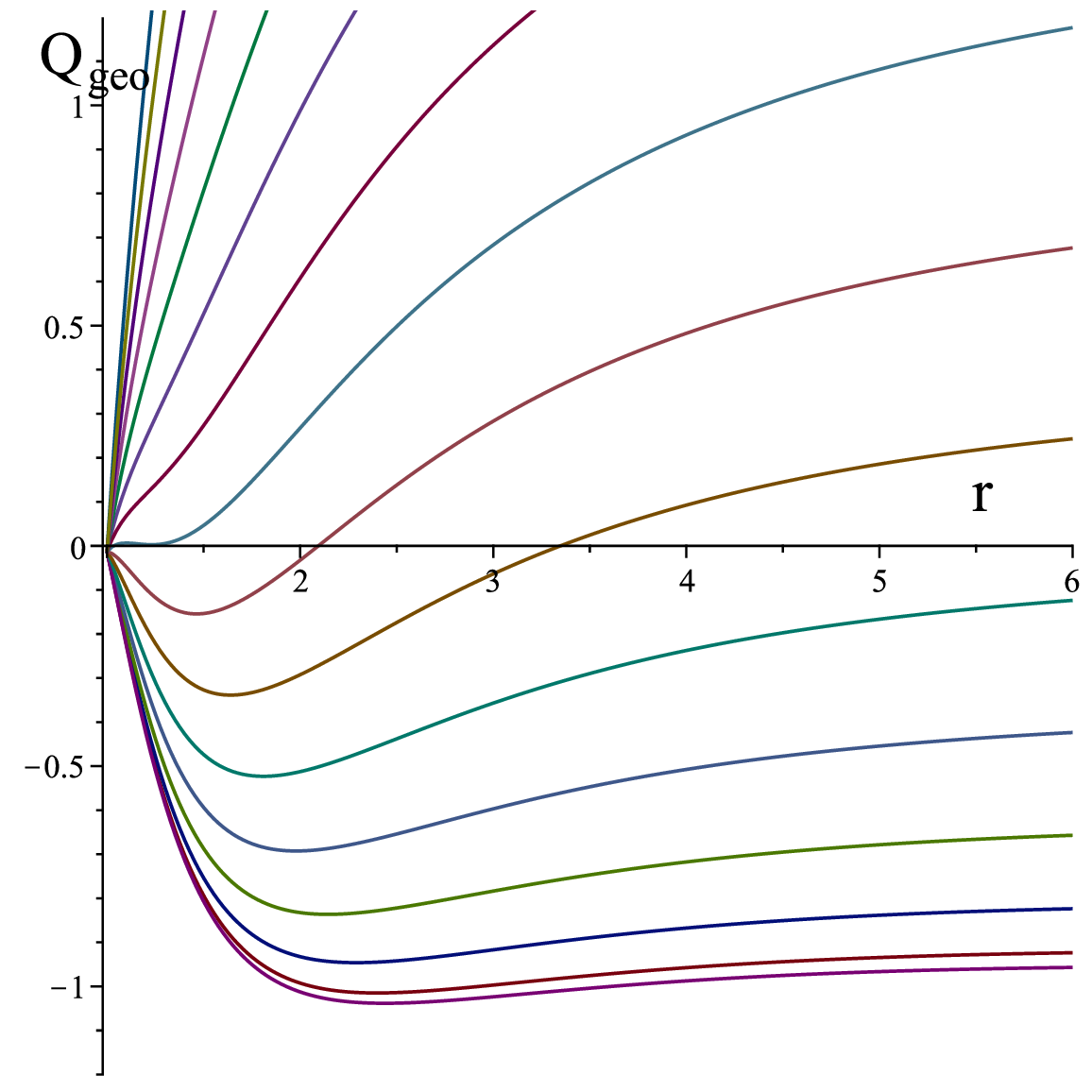}&\qquad
\includegraphics[scale=0.30]{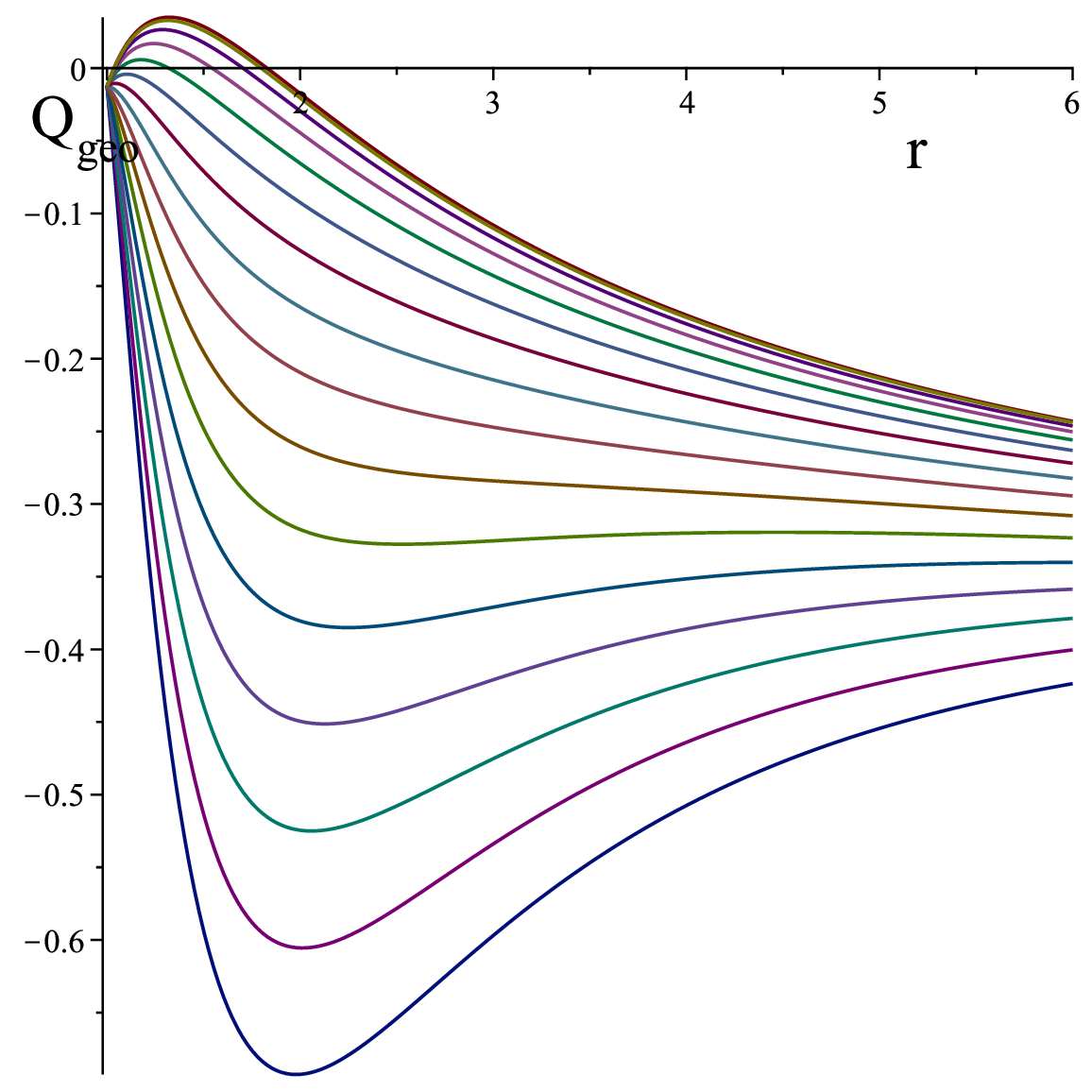}\cr
(a) &\qquad (b) \cr
\includegraphics[scale=0.30]{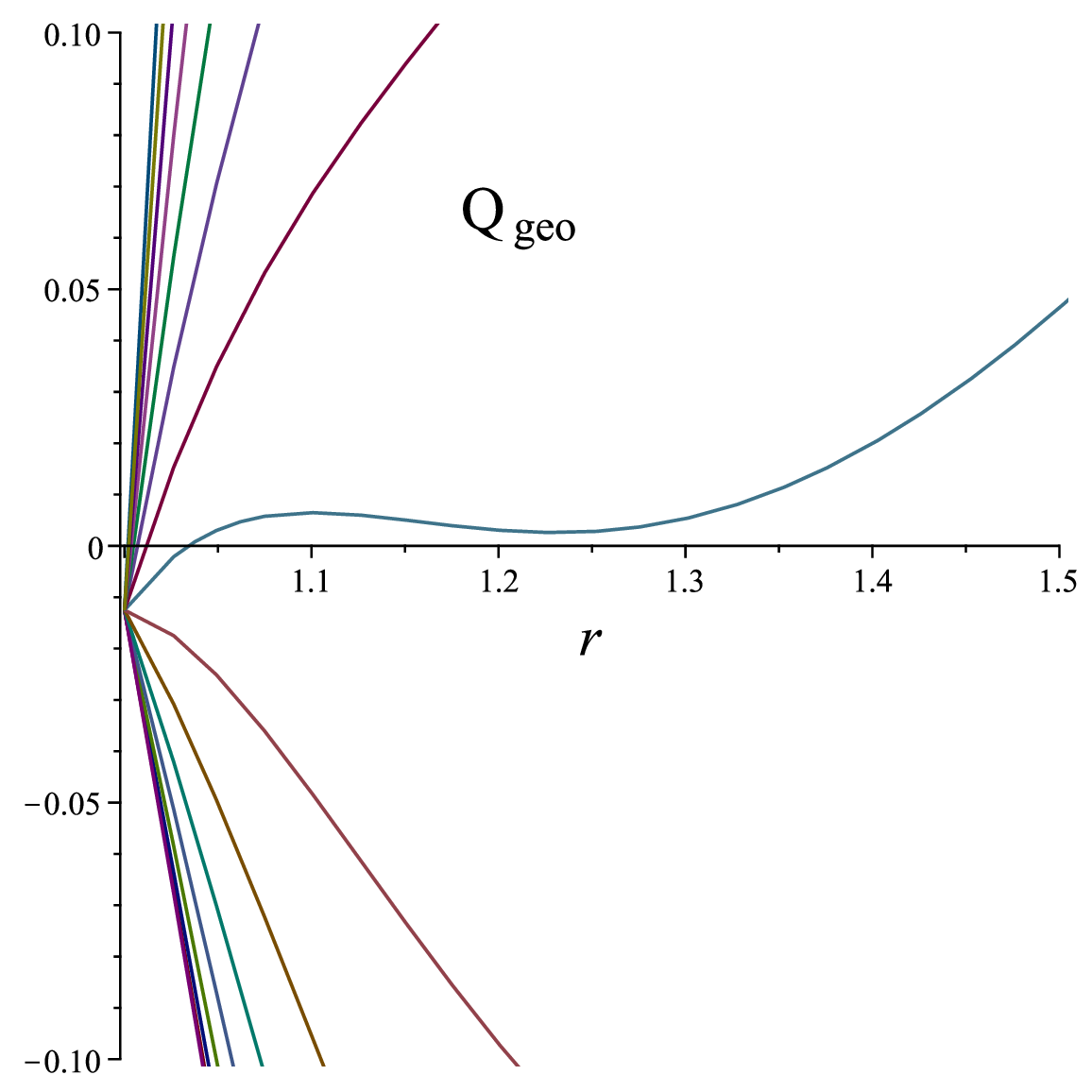}&\qquad
\includegraphics[scale=0.30]{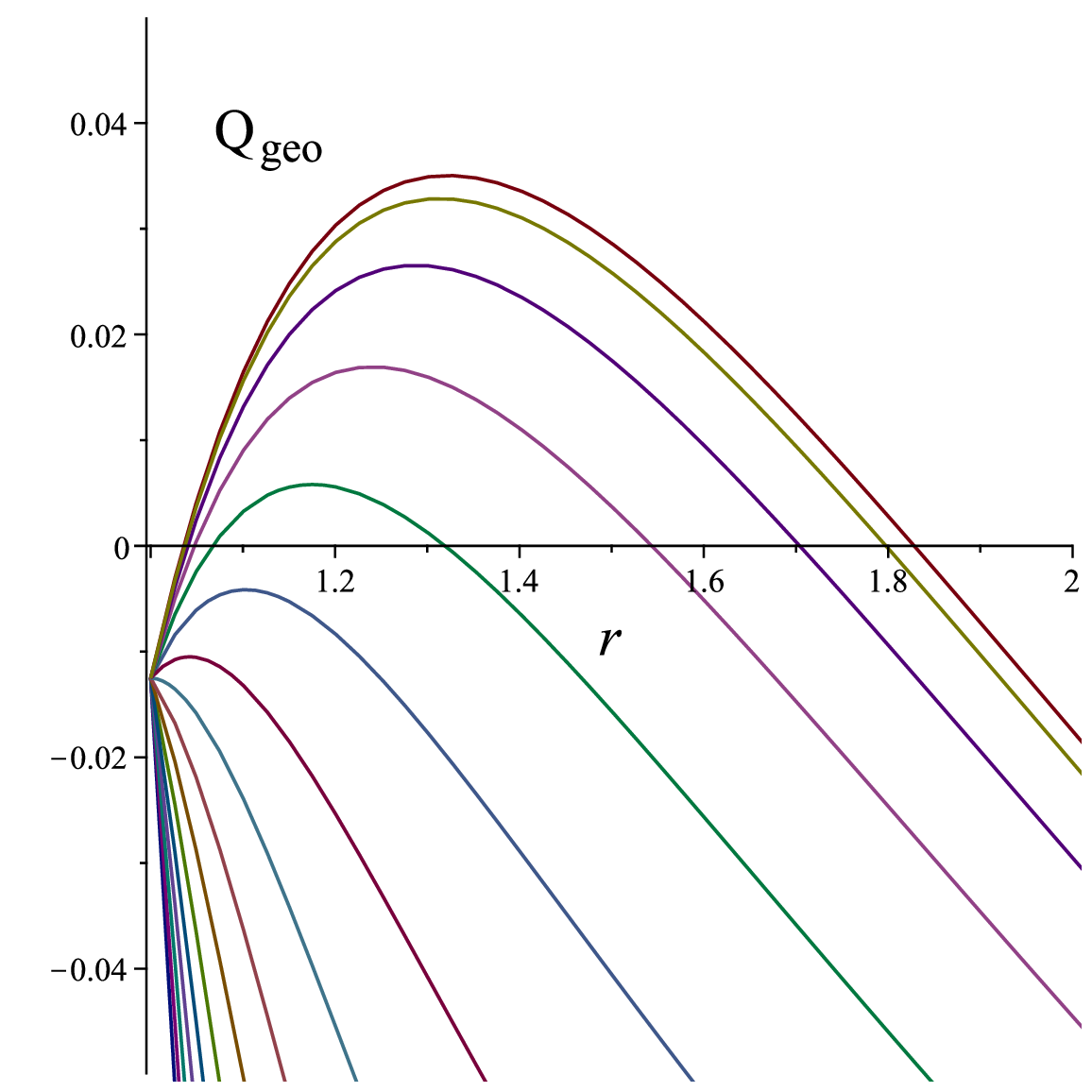}\cr
(c) &\qquad (d) \cr
\end{array}
\]
\caption{\label{fig12} $Q_{\rm geo}$ as given in Eq. \eqref{veffts} is plotted as a function of $r$ with the following choice of parameters $\mu=1,r_b=1,r_s=0.8,p=1/4$. In   panel (a) we have chosen $J=3$ and varied the energy according to the relation $E=1+k/10$ for $k$ varying between $[-20,20]$  with a step of $2$ (curves ordered from below). 
In  panel (b) we have chosen $E=0.8$ and varied the angular momentum according to the relation $J=1+k/10$ for $k$ varying between $[-20,20]$  with a step of $2$ (curves ordered from above). Panel (c) and (d) are detailed views of panels (a) and (b), respectively.
}
\end{figure*}

Depending on the value of $E$ and $J$, and thus of the impact parameter $b=J/\sqrt{E^2-\mu^2}$, 
one can have open (unbound `hyperbolic') or closed (rather `elliptic' bound with $r_m< r<r_M$) orbits that become critical for finely tuned choices of $E=E_c$ and $K=K_c$ so as to get a circular geodesics with $r=r_c$.
 
Thanks to spherical symmetry, also non-critical geodesics are planar and without any loss of generality, one can focus again on the equatorial plane $\theta=\pi/2$, with  $\dot{\theta}=P_\theta=0$ and $K^2=J^2$. 

While for critical geodesics $r=r_c$, the last of \eqref{conjcan} implies $\dot\phi = J/r_c^2=\omega_c$, for a non-critical geodesic $\dot{\phi}={J\over r^2}$ is not a constant. Taking the ratio between $\dot{r}$ and $\dot{\phi}$ and integrating one finds 
\be
\phi-\phi_0=J\int_{r_0}^r{dr'\over (r'-r_s)(r'-r_b)\sqrt{Q_r(r')}}\,,
\ee
where the radial integral can be written in the form 
\bea
I_r&=&\int{dr\over (r-r_s)(r-r_b)\sqrt{Q_r(r)}} \nonumber\\
&=& \int {dr\over \sqrt{\prod_{i=1}^4(r-r_i)}}\,,
\eea
with $r_i$ the (ordered) solutions of the quartic equation (see Eq. \eqref{prpq} above)
\bea
&&r^2[r(r-r_b)E^2-r(r-r_s){p}^2-(r-r_b)(r-r_s)\mu^2]\nonumber\\
&&-(r-r_b)(r-r_s)K^2=0\,,
\eea
and expressed in terms of incomplete elliptic integral of the first kind 
\bea
\label{elliptic}
I_r&=&{2\over \sqrt{r_{13}r_{24}}}\mathcal{K}\Big[\arcsin\left({\sqrt{(r-r_1)r_{24}\over(r-r_2)r_{14}}}\right),{r_{23}r_{14}\over r_{13}r_{24}}\Big]\,,\nonumber\\
\eea
where $r_{ij}=r_i-r_j$. The positive real roots represent potential turning points of the radial motion, where $\dot{r}\sim P_r = 0$. Ordering the turning points according to
$r_1\ge r_2 ...$, an unbound orbit corresponds to radial motion with $r\ge r_1$.
Bound orbits appear either when $r_1>r_2\ge r\ge r_3\ge r_b$ (internal to the `photon-ring') 
or when $r_1>r>r_2 \ge r_b$ (external to the `photon-ring'). The latter represents an `elliptic' deformation of the ISCO. We will sketch how to modify our later analysis to this potentially very interesting kind of non-critical orbits.

Let us specialize the above results to the case of timelike (equatorial) circular geodesics (with $P_r=P_\theta=P_y=0$, $r=r_0$ and $\theta=\pi/2$ and hence $K=J$) which exist at a constant radius $r=r_0$, such that
\beq
\label{pot_eff_eq}
\frac{E^2}{\mu^2}-1-\frac{1}{\mu^2}V_{\rm eff}(r_0)=0\,.
\eeq
Let us define energy and angular  momentum per unit mass 
\beq
\hat E=\frac{E}{\mu}\,,\qquad \hat J=\frac{J}{\mu}\,.
\eeq
Eq. \eqref{pot_eff_eq} factorizes as
\beq
\left(1-\frac{r_b}{r_0}\right)\left[- \hat E^2+f_s(r_0)\left(1 +\frac{\hat J^2}{r_0^2} \right)\right]=0\,,
\eeq
and admits the solution
\beq 
r_0=r_b\,,
\eeq
which is either unstable (for $r_b>3r_s/2$) or stable but internal to the photon-sphere (for $r_b<3r_s/2$). The other solution corresponds to
\beq
\label{normaliz}
- \frac{\hat E^2}{f_s(r_0)}+1 +\frac{\hat J^2}{r_0^2}=0\,,
\eeq
which is exactly the same relation valid in the Schwarzschild spacetime.
Eq. \eqref{normaliz} can be also seen as 
the normalization condition for the four velocity $U$ related to the momentum by $P^\mu=\mu U^\mu$. 
In fact, for circular orbits it is convenient to write $U=\Gamma (\partial_t +\Omega \partial_\phi)$, where $\Omega=\frac{d\phi}{dt}$ denotes the corresponding angular velocity. 
Moreover, the Killing symmetries of the spacetime imply $U_t=-{\hat E}$, $U_\phi={\hat J}$, that is
\beq
U^t=\frac{dt}{d\tau}=\Gamma=\frac{\hat E}{f_s(r_0)}\,,\qquad U^\phi=\frac{d\phi}{d\tau}=\Gamma \Omega=\frac{\hat J}{r^2_0}\,.
\eeq
The condition $U\cdot U=-1$ then reads
\beq
g_{tt}(U^t)^2+g_{\phi\phi}(U^\phi)^2=-1\,,
\eeq
and reduces to Eq. \eqref{normaliz}. Moreover,
\beq
\Gamma=\frac{1}{\sqrt{1-\frac{r_s}{r_0}-\Omega^2 r^2_0}}\,,
\eeq
as it follows by replacing $\hat E=\Gamma f_s(r_0)$ and $\hat J=\Gamma r_0^2 \Omega$ in Eq. \eqref{normaliz}.
Finally, the critical geodesic condition  
\beq
\frac{d}{dr}V_{\rm eff}(r)\bigg|_{r=r_0}=0\,,
\eeq
with
\beq
\frac{1}{\mu^2}V_{\rm eff}(r)=\frac{r_b (\hat E^2-1)}{r}- \frac{r_s}{r}f_b(r)+\frac{\hat J^2}{r^2}f_s(r)f_b(r)\,.
\eeq
implies  $\Omega =\sqrt{ \frac{r_s}{2r_0^3} }$.
Summarizing ISCO (innermost stable circular geodesics, yet external to the photon-sphere) correspond to $r=r_0=r_{c3} > r_{c2}>r_b $ with angular velocity, relativistic factor, energy and angular momentum given by
\beq
\label{Omega_circ}
\Omega=\sqrt{\frac{r_s}{2r_0^3}} \,,\qquad \Gamma =\frac{1}{\sqrt{1-\frac{3r_s}{2r_0}}}\,,
\eeq
and
\beq
\label{E_circ}
\hat E=\frac{1-\frac{r_s}{r_0}}{\sqrt{1-\frac{3r_s}{2r_0}}}\,,\qquad \frac{\hat J}{M}=\frac{1}{\sqrt{\frac{r_s}{2r_0}\left(1-\frac{3r_s}{2r_0}\right)}}\,.
\eeq
Later on, we will consider massless scalar wave emission from a massive probe (with $\mu\ll M_{\rm TS}$). To this end, we will now proceed by analyzing the source-free massless scalar field equation.

\section{Nongeodesic motion: acceleration effects due to superposed  test  fields}

Suppose that a test field, described by an energy-momentum tensor $T^{\mu\nu}_{\rm test}$, is superposed to the TS spacetime in the sense that
\beq
\nabla_\mu T^{\mu\nu}_{\rm test}=0\,,
\eeq
namely it satisfies a covariant conservation condition with respect to the background metric.
For example, $T^{\mu\nu}_{\rm test}$ can be associated either with a radiation field or a perfect fluid and can be nonzero in some spacetime region (bounded or unbounded). A test particle moving on the TS spacetime can interact with $T^{\mu\nu}_{\rm test}$ which is then responsible for acceleration effects.
If the five-velocity of the particle is denoted by $U=U^\lambda\partial_\lambda$, then the acceleration $a(U)^\lambda=U^\mu \nabla_\mu U^\lambda$ of the particle satisfies an acceleration-equal-force equation of the type
\beq
\mu a(U)^\lambda=\sigma P(U)^\lambda{}_\mu T^{\mu\nu}_{\rm test}U_\nu\,,
\eeq
where $P(U)^\lambda{}_\mu=\delta^\lambda{}_\mu+U^\lambda U_\mu$ projects orthogonally to $U$.
Equivalently,
\beq
\label{aU_PR}
a(U)^\alpha=\tilde \sigma P(U)^\alpha{}_\beta T^{\beta\nu}_{\rm test}U_\nu\,,\quad \tilde \sigma=\frac{\sigma}{\mu}\,.
\eeq
Clearly, in this simplified picture we are ignoring back-reaction effects.
In the literature this approach is known as Poynting-Robertson approach\cite{Poy1903,Rob1937} (see also Refs. \cite{Bini:2008vk,Bini:2011cll,Bini:2012wot,Bini:2012ncd} for a recent review). In a hierarchy of length scales involved, the length scale associated with the background is assumed here to be much larger than any other length scale associated with both the particles or the superposed fields. Clearly, this is also a limitation for the present study and leads to the necessity to study backreaction effects.  

We will see below two explicit examples, concerning a (test)  radiation field and a (test) dust field which we assume in both cases to be directed radially (inward, outward). Generalizations of the latter assumption (including for example rotating photon or dust fields) can be easily developed, and we will not proceed further in this direction.

\subsection{Particle scattering by a test radiation field}

A more realistic situation would be a TS surrounded by  a radiation fields. In absence of exact solutions of this type we follow here the so-called Poynting-Robertson model\cite{Poy1903,Rob1937}, briefly outlined below.

Let $U$ be the five velocity of a test particle moving on the equatorial plane
\beq
U=\gamma [e_0+{{v}} {\mathbf n}]\,,
\eeq
where $\gamma=(1-{{v}}^2)^{-1/2}$ is the Lorentz factor and
\beq
n=\sin\alpha \cos\beta e_r +\sin \alpha \sin \beta e_\phi+\cos \alpha e_y
\eeq
is a  unit spatial vector in five dimension, i.e., including the Kaluza-Klein direction $e_y$.

Imagine a test radiation field superposed to the TS background and described by the following energy-momentum tensor
\beq
T^{\mu\nu}=\Phi k^\mu k^\nu\,,
\eeq
with $k$ a null vector in the $t-r$ plane, 
\beq
k=\partial_t +f_s(r)\sqrt{f_b(r)}\partial_r\,,
\eeq
and $\Phi=\Phi(r)$ a (radial) flux factor.
The divergence-free condition for $T$, 
$\nabla_\mu T^{\mu\nu}=0,$
determines completely   $\Phi$
\beq
\Phi= \frac{1}{f_s(r)^2\sqrt{f_b(r)}}\,.
\eeq
The radiation field contributes to the acceleration of test particles because of the associated force
\beq
F_{\rm rad}^\lambda = \sigma P(U)^\lambda{}_\mu T^{\mu\nu}U_\nu\,,
\eeq
where $\sigma$ is a coefficient representing the efficiency of the absorption-re-emission of radiation by the particle, and $P(U)=g+U\otimes U$ projects orthogonally to $U$.
We will study the motion of a mass $\mu$ test particle according to the law
\beq
\label{acc_eq}
\mu a(U)=F_{\rm rad}\,,
\eeq
with $a(U)=\nabla_UU$ the acceleration of the type
\beq
a(U)^\lambda=\tilde \sigma P(U)^\lambda{}_\mu T^{\mu\nu}U_\nu\,,\qquad \tilde \sigma=\frac{\sigma}{\mu}\,.
\eeq
The equations of motion read
\bea
\frac{d{{v}}}{d\tau}&=& -\frac{r_s}{2r^2\gamma}\sqrt{\frac{f_b(r)}{f_s(r)}} \sin \alpha \cos\beta\nonumber\\ 
&-&\tilde \sigma \frac{\sin \alpha \cos \beta(1+ {{v}}^2) -(1+\sin^2\alpha \cos^2\beta){{v}}}{r^2 f_s(r)\sqrt{f_b(r)}}\,, \nonumber\\
\frac{d\alpha}{d\tau}&=&-\frac{\gamma [r_sf_b(r)-r_b f_s(r){{v}}^2]}{2{{v}} r^2 \sqrt{f_s(r)f_b(r)}}\nonumber\\
&+&\tilde \sigma  \frac{({{v}} \sin \alpha \cos \beta-1)\cos\alpha \cos\beta}{r^2 {{v}} f_s(r)\sqrt{f_b(r)}}\,, \nonumber\\
\frac{d\beta}{d\tau}&=& \frac{\gamma \sin \beta [r_sf_b(r)-f_s(r){{v}}^2(r_b+\sin^2\alpha (2r-3r_b))]}{ 2{{v}} r^2 \sqrt{f_b(r) f_s(r) }}\nonumber\\
&-&\tilde\sigma \frac{
\sin \beta ({{v}} \sin \alpha \cos\beta-1)}{{{v}} r \sin \alpha f_s(r)\sqrt{f_b(r)}  }\,,
\eea
together with the equatorial plane condition $\theta=\pi/2$ and 
\bea
\label{equat_pl_conds}
\frac{dt}{d\tau}&=& \frac{\gamma}{\sqrt{f_s(r)}}\,,\nonumber\\
\frac{dr}{d\tau}&=&  \gamma{{v}}\sin\alpha \cos \beta \sqrt{f_s(r)} \sqrt{f_b(r)}\,,\nonumber\\
\frac{d\phi}{d\tau}&=&  \frac{\gamma v}{r}\sin\alpha \sin \beta \,, \nonumber\\
\frac{dy}{d\tau}&=& \frac{\gamma {{v}} \cos \alpha}{\sqrt{f_b(r)}}\,.
\eea
The geodesic case is recovered when $\tilde \sigma=0$.
The main feature of the  motion \eqref{acc_eq} is that the accelerated particle, gravitationally attracted by the TS, loses energy during its motion, and at a certain point it can  stop. If, instead of braking effects, one assumes an outward acceleration due to the radiation field, attraction by the TS and outward acceleration compete: the particle can be either attracted towards the `cap'  or pushed away. 
Examples of numerical integration of the orbits are given below.

\begin{figure*}[h]
\[
\begin{array}{cc}
\includegraphics[scale=0.35]{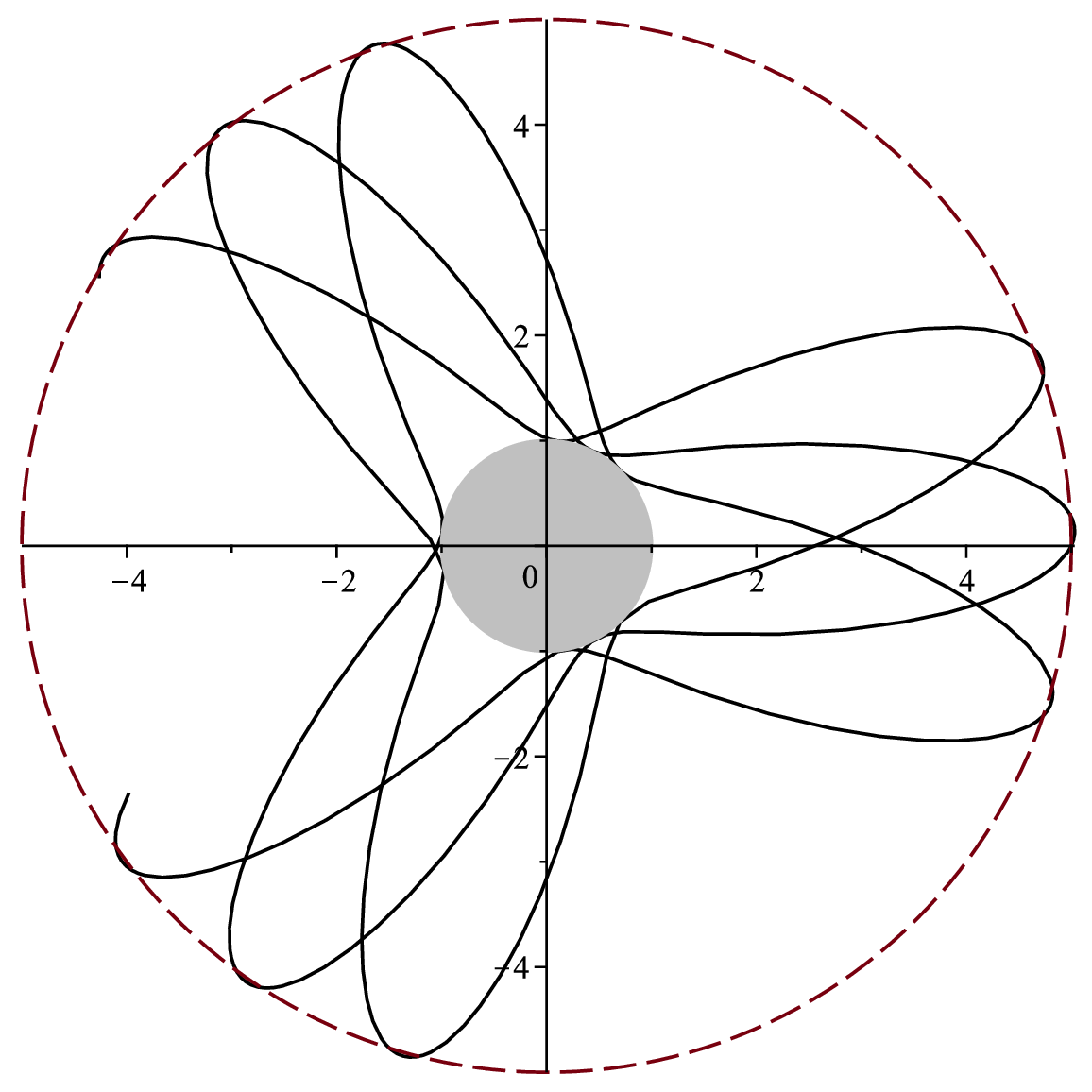}& 
\includegraphics[scale=0.35]{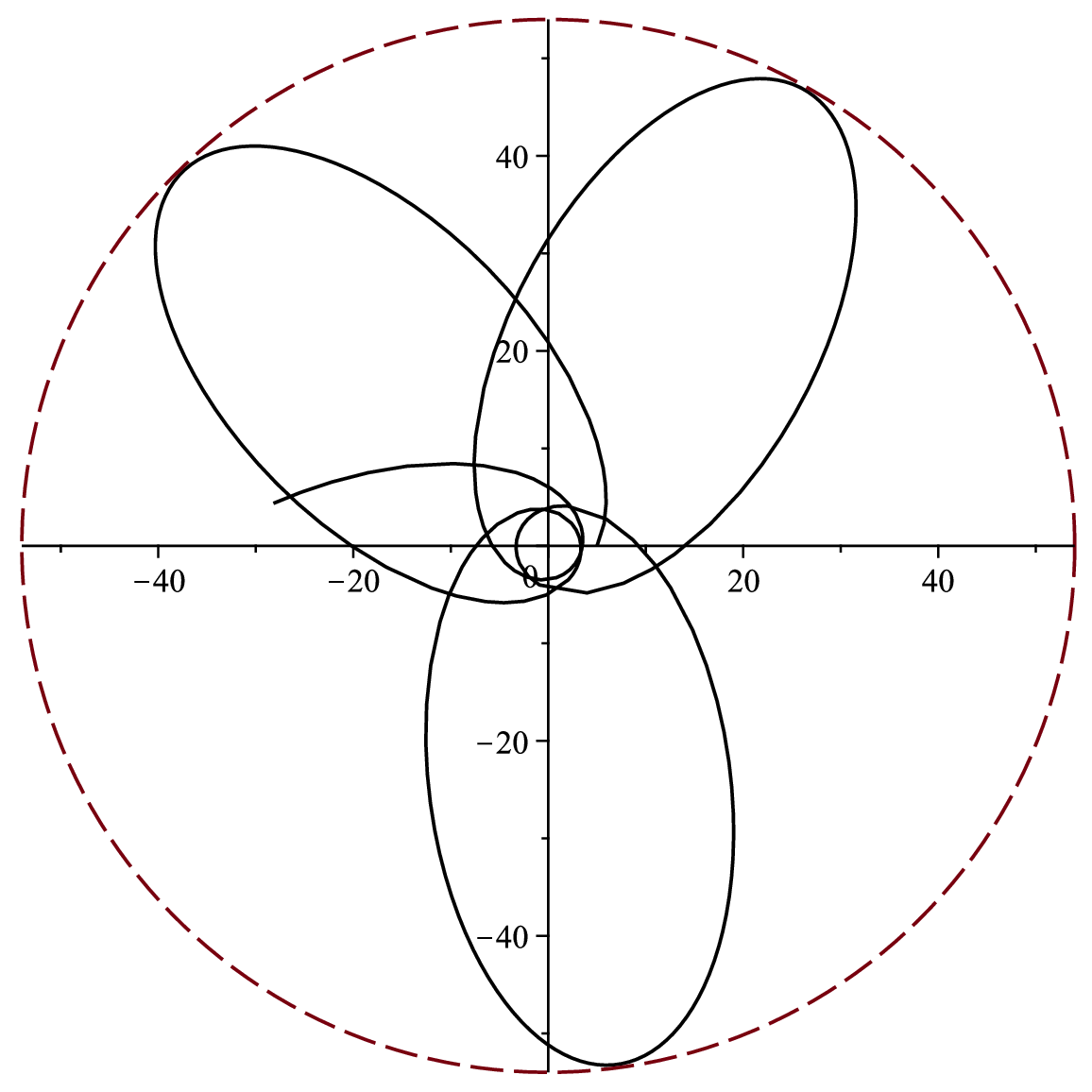}\\
(a) & (b)\\
\end{array}
\]
\caption{\label{fig:geo1} Panel (a): Section $X-Y$ ($X=r(\tau)\cos(\phi(\tau)$, $Y=r(\tau)\sin(\phi(\tau)$) of the geodesic motion ($\tilde \sigma=0$) in the bound case, with background parameters $r_s=0.8$ and $r_b=1$ (filled circle, grey online).  
Initial conditions are chosen so that $\alpha(0)=\pi/4$, $\beta(0)=\pi/3$, $v(0)=0.1$, $r(0)=5$, $y(0)=0.1$, $\phi(0)=0=t(0)$. The motion is contained in a corona with inner radius $r_{\rm int}=r_b=1$ and outer radius $r_{\rm ext}=5$.
Panel (b): The background parameters and initial conditions as in Panel (a), except for $v(0)=0.5$.
}
\end{figure*}

\begin{figure*}[h]
\[
\begin{array}{cc}
\includegraphics[scale=0.35]{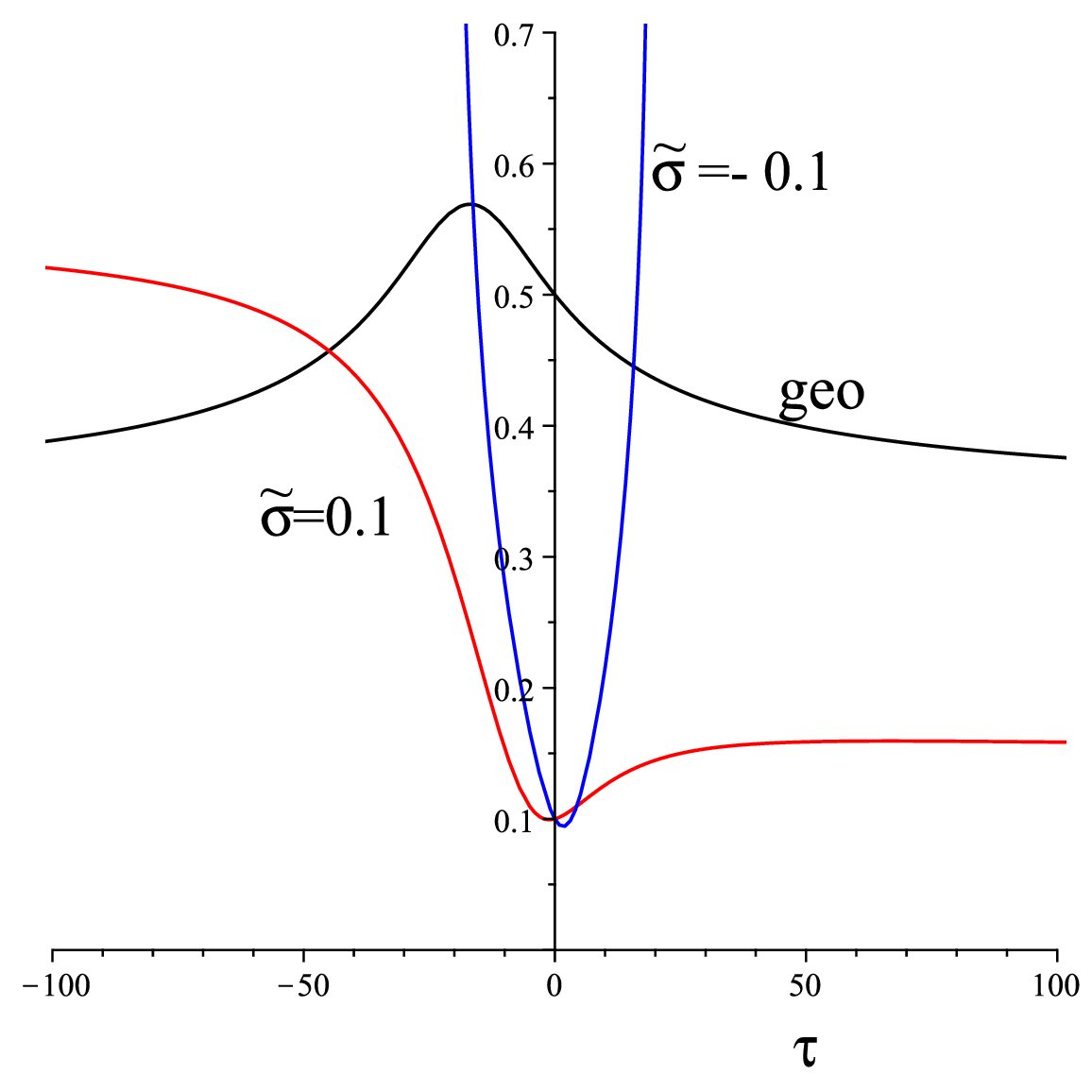}& 
\includegraphics[scale=0.35]{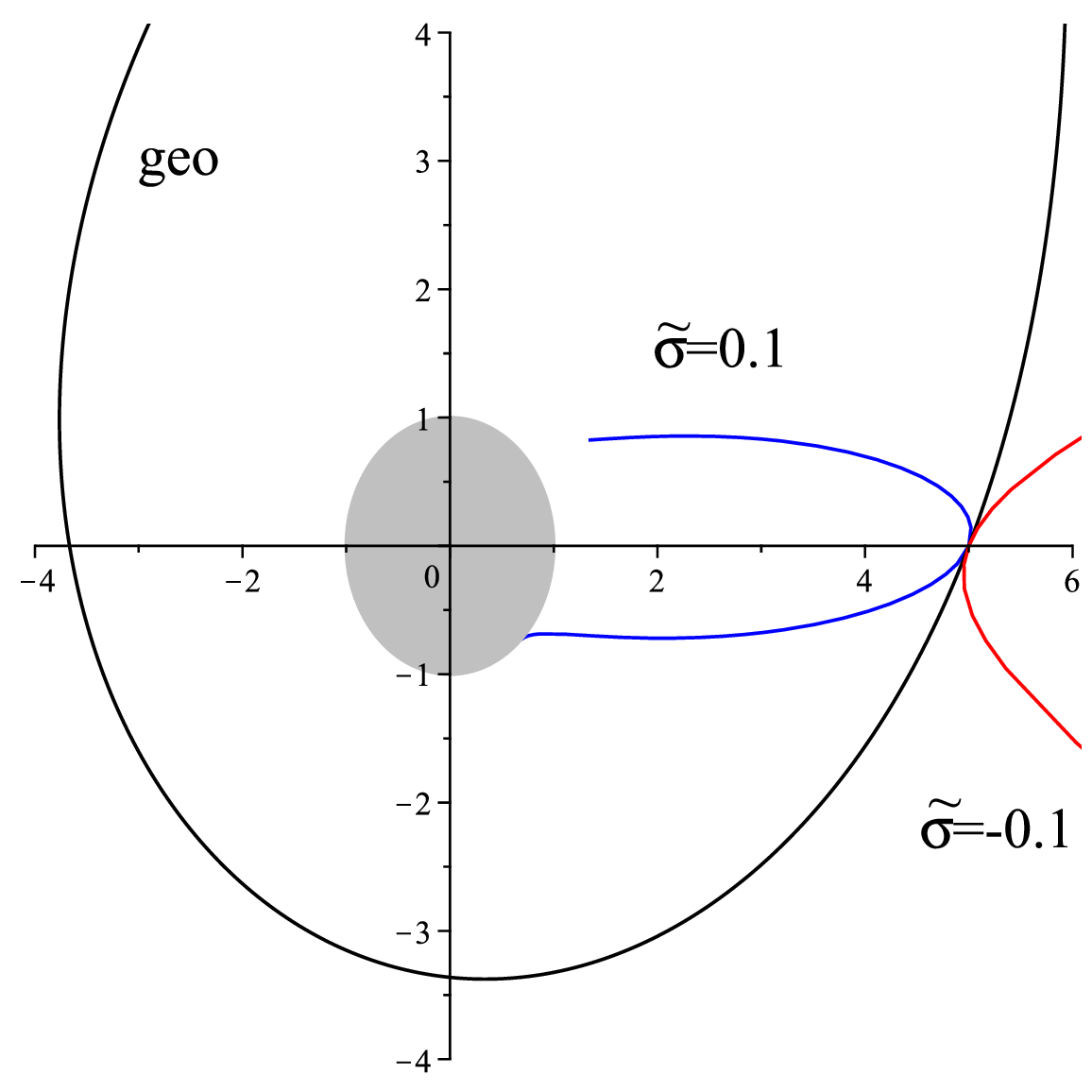}\\
(a) & (b)\\
\end{array}
\]
\caption{\label{fig:acc1} Panel (a): Behavior of the velocity $v(\tau)$ in the cases $\tilde \sigma=[-0.1,0,0.1]$ with the choice of background parameters as in Fig. \ref{fig:geo1}, i.e.,  $r_s=0.8$ and $r_b=1$  and  
initial conditions  $\alpha(0)=\pi/4$, $\beta(0)=\pi/3$, $v(0)=0.1$, $r(0)=5$, $y(0)=0.1$, $\phi(0)=0=t(0)$.
Panel (b): 
Section $X-Y$ ($X=r(\tau)\cos(\phi(\tau)$, $Y=r(\tau)\sin(\phi(\tau)$) of the motion in the cases $\tilde \sigma=[-0.1,0,0.1]$ with same choice of parameters and initial conditions as in Panel (a).
}
\end{figure*}

There exist special cases which can be examined with more detail. For example, the case
\beq
\alpha=\frac{\pi}{2}\,,\qquad \beta=0\,,
\eeq
implying $n=e_r$ (radial flow of photons on the equatorial plane and on the $y=0$ surface) and
\bea
\frac{d v}{d\tau}&=& -\tilde \sigma \frac{(1-v)^2}{(r-r_s)\sqrt{r}\sqrt{r-r_b}}-\frac{1}{2\gamma}\frac{r_s}{r^2}\sqrt{\frac{r-r_b}{r-r_s}}\nonumber\\
\frac{dt}{d\tau}&=& \frac{\gamma}{\sqrt{1-\frac{r_s}{r}}}\,,\nonumber\\
\frac{dr}{d\tau}&=& \frac{\gamma v}{r} \sqrt{(r-r_s)(r-r_b)}\,,
\eea
as well as 
\beq
\frac{d\phi}{d\tau}=0=\frac{dy}{d\tau}\,.
\eeq
Here $v$ will depend on $\tilde\sigma$ and hence also $r$. In this particular case it can be convenient to use $r$ as a parameter so that the problem is reduced to the single equation
\beq
\frac{dv}{dr}=-\tilde \sigma \frac{(1-v)^2\sqrt{1-v^2}\sqrt{r}}{v (r-r_s)^{3/2}(r-r_b)}-\frac12 \frac{(1-v^2)}{v(r-r_s)} \frac{r_s}{r}\,,
\eeq
which, for example,  can be solved perturbatively in $\tilde \sigma$ leading to a solution for $v$ as a first-order correction to the geodetic value.

\subsection{Particle scattering by a test dust field}

Another realistic situation would be a TS surrounded by a matter  field which we can examine again \`a la Poynting-Robertson \cite{Poy1903,Rob1937}, as briefly outlined below.

Let $U$ be the five velocity of a test particle moving on the equatorial plane
\beq
U=\gamma [e_0+{{v}} {\mathbf n}]\,,
\eeq
where $\gamma=(1-{{v}}^2)^{-1/2}$ is the Lorentz factor and
\beq
n=\sin\alpha \cos\beta e_r +\sin \alpha \sin \beta e_\phi+\cos \alpha e_y
\eeq
is a  unit spatial vector in five dimension, i.e., including the Kaluza-Klein direction $e_y$.

Imagine a test matter (dust) field superposed to the TS background and described by the following energy-momentum tensor
\beq
T^{\mu\nu}=\Phi u^\mu u^\nu\,,\qquad u\cdot u=-1\,,
\eeq
with $u$ a timelike vector (one-form, denoted by the symbol $\flat$) field (parametrized, in this case,  by the energy parameter $E$) in the $t-r$ plane, 
\bea
u&=&\frac{E}{f_s(r)}\partial_t+\sqrt{f_b(r)}\sqrt{E^2-f_s(r)}\partial_r\,,\nonumber\\
u^\flat&=&-Edt+\frac{ \sqrt{E^2-f_s(r)}}{f_s(r)  \sqrt{f_b(r)}}dr\,,
\eea
and $\Phi=\Phi(r)$ a (radial) flux factor.
The conservation condition for $T$, 
$\nabla_\mu T^{\mu\nu}=0,$
determines completely   $\Phi$
\beq
\Phi= \frac{1}{r^2\sqrt{f_b(r)}\sqrt{E^2-f_s(r)}}\,.  
\eeq
The dust field contributes to the acceleration of test particles because of the associated force
\beq
F_{\rm rad}^\lambda = \sigma P(U)^\lambda{}_\mu T^{\mu\nu}U_\nu\,,
\eeq
where $\sigma$ is a coefficient representing the intensity of the friction, and $P(U)=g+U\otimes U$ projects orthogonally to $U$.
We will study the motion of a mass $\mu$ test particle according to the law
\beq
\label{acc_eq1}
\mu a(U)=F_{\rm rad}\,,
\eeq
with $a(U)=\nabla_UU$ the acceleration of the type
\beq
a(U)^\lambda=\tilde \sigma P(U)^\lambda{}_\mu T^{\mu\nu}U_\nu\,,\qquad \tilde \sigma=\frac{\sigma}{\mu}\,.
\eeq
The equations of motion read
\bea
\frac{d{{v}}}{d\tau}&=& -\frac{r_s}{2r^2\gamma}\sqrt{\frac{f_b(r)}{f_s(r)}} \sin \alpha \cos\beta\nonumber\\ 
&+&\tilde \sigma\left[ v\sin^2\alpha\cos^2\beta \frac{\sqrt{E^2-f_s(r)}}{ r^2 f_s(r) \sqrt{f_b(r)}}\right.\nonumber\\
&& -\sin(\alpha)\cos\beta\frac{ E (v^2+1)}{ r^2 f_s(r) \sqrt{f_b(r)}}\nonumber\\
&&\left. +\frac{E^2 v}{ r^2 f_s(r) \sqrt{f_b(r)}\sqrt{E^2-f_s(r)}} \right] \,, \nonumber\\
\frac{d\alpha}{d\tau}&=&-\frac{\gamma [r_sf_b(r)-r_b f_s(r){{v}}^2]}{2{{v}} r^2 \sqrt{f_s(r)f_b(r)}}\nonumber\\
&+&\tilde \sigma\left[\frac{\cos(\alpha)\sin(\alpha)\cos^2(\beta)\sqrt{E^2-f_s(r)}}{ r^2 \sqrt{f_b(r)} f_s(r)}\right.\nonumber\\
&&\left.-\frac{E\cos(\beta)\cos(\alpha)}{r^2 v  \sqrt{f_b(r)} f_s(r)} \right] \,, \nonumber\\
\frac{d\beta}{d\tau}&=& \frac{\gamma \sin \beta [r_sf_b(r)-f_s(r){{v}}^2(r_b+\sin^2\alpha (2r-3r_b))]}{ 2{{v}} r^2 \sqrt{f_b(r)} f_s(r)  }\nonumber\\
&+&\tilde \sigma\left[\frac{E \sin(\beta)}{r^2\sin(\alpha) v \sqrt{f_b(r)} f_s(r)}\right.\nonumber\\ 
&&\left.-\frac{\sin(\beta)\cos(\beta)\sqrt{E^2-f_s(r)}}{r^2\sqrt{f_b(r)} f_s(r) } \right] \,,
\eea
together with the equatorial plane condition $\theta=\pi/2$ and the Eqs. \eqref{equat_pl_conds} as in the previous case,  
\bea
\frac{dt}{d\tau}&=& \frac{\gamma}{\sqrt{f_s(r)}}\,,\nonumber\\
\frac{dr}{d\tau}&=&  \gamma{{v}}\sin\alpha \cos \beta \sqrt{f_s(r)} \sqrt{f_b(r)}\,,\nonumber\\
\frac{d\phi}{d\tau}&=&  \frac{\gamma v}{r}\sin\alpha \sin \beta \,, \nonumber\\
\frac{dy}{d\tau}&=& \frac{\gamma {{v}} \cos \alpha}{\sqrt{f_b(r)}}\,.
\eea
As before, the geodesic case is recovered when $\tilde \sigma=0$.
and the special case $\alpha=\frac{\pi}{2}$ and $\beta=0$ is much simplified
\bea
\frac{d{{v}}}{d\tau}&=& -\frac{r_s}{2r^2\gamma}\sqrt{\frac{f_b(r)}{f_s(r)}} \nonumber\\ 
&+&\frac{\tilde \sigma}{r^2 f_s(r) \sqrt{f_b(r)}}\left[ v  \sqrt{E^2-f_s(r)} 
 - E (v^2+1) \right.\nonumber\\
&&\left. +\frac{E^2 v}{ \sqrt{E^2-f_s(r)}} \right] \,, \nonumber\\
v^2&=&\frac{r_sf_b(r)}{r_b f_s(r) } \,,
\eea
and can be studied analytically. See e.g., Refs. \cite{Bini:2008vk,Bini:2011cll,Bini:2012wot,Bini:2012ncd}.

\section{Source-free massless scalar field}

In the TS background, a massless neutral scalar perturbation satisfies the equation
\beq
\Box\phi=g^{\mu\nu}\nabla_\mu \partial_\nu \phi=0\,.
\eeq
Separating the angular variables and Fourier-transforming  (FT) over $t$ (integral) and $y$ (series) yield
\beq
\phi=\sum_{\ell,m} \sum_{{p}}\int \frac{d\omega}{2\pi} e^{-i\omega t}\frac{e^{i{p} y}}{2\pi} Y_{\ell m}(\theta,\phi)R_{(\ell,m,\omega,{p})}(r)\,,
\eeq
with $Y_{\ell m}(\theta,\phi)=e^{im\phi}P_{\ell m}(\theta)$ the standard spherical harmonics and $P_{\ell m}(\theta)$ the associated Legendre functions that satisfy
\beq
\frac{d^2P_{\ell m}}{d\theta^2}+\cot(\theta)\frac{dP_{\ell m}}{d\theta} +\left(\ell (\ell+1)-\frac{m^2}{\sin^2(\theta)}\right) P_{\ell m}(\theta)=0\,,
\eeq
The radial equation satisfied by $R_{(\ell,m,\omega,{p})}(r)\equiv R(r)$ (abbreviated for simplicity) reduces to
\begin{widetext}
\bea
\label{tsw1}
&&\frac{d^2R}{dr^2}+\left(\frac{1}{r-r_b}+\frac{1}{r-r_s}\right)\frac{dR}{dr}
+ \frac{r^3}{(r-r_s)(r-r_b)}\left(\frac{\omega^2}{r-r_s}-\frac{{p}^2}{r-r_b}- \frac{\ell(\ell+1)}{r^3} \right) R(r)=0\,.\nonumber\\
\eea
\end{widetext}
We will use the notation $L=\ell(\ell+1)$ when convenient.
Eq. \eqref{tsw1} admits a general solution of the form
\bea
R(r)&=&e^{i \tilde\omega r } (r-r_b)^{ {\lambda}}[C_1 (r-r_s)^{-{\kappa}}H_1(r)\nonumber\\
&+& C_2(r-r_s)^{+{\kappa} }H_2(r)]\,,
\eea
where  $C_{1}$ and $C_2$ are integration constants, $H_{1,2}(r)$ are Confluent Heun Functions (CHFs), see below, and   
\bea
\label{om_lam_kappa_defs}
\tilde\omega&=&\sqrt{\omega^2-{p}^2}\,,\nonumber\\
{\lambda}&=&\frac{ |p| r_b^{3/2}}{(r_b-r_s)^{1/2}}\,,\nonumber\\
{\kappa}&=&\frac{\omega r_s^{3/2}}{(r_b-r_s)^{1/2}}\,.
\eea
We consider $\omega>|{p}|$ (outgoing waves when $r \to \infty$, regularity in $r=r_b$).
Eq. \eqref{tsw1} is the main radial equation which we will consider in detail below in the special case ${p}=0$ ($\tilde\omega=\omega$),
\bea
\label{tsw1pyzero}
&&\frac{d^2R}{dr^2}+\left(\frac{1}{r-r_b}+\frac{1}{r-r_s}\right)\frac{dR}{dr}\nonumber\\
&&+\frac{r^3}{(r-r_s)(r-r_b)}\left(\frac{\omega^2}{r-r_s}- \frac{\ell(\ell+1)}{r^3} \right) R(r)=0\,.\nonumber\\
\eea
Replacing 
\beq
R(r)=e^{i\omega r} (r-r_s)^{-{\kappa}}H(r)
\eeq
in Eq. \eqref{tsw1pyzero}
the equation satisfied by $H(r)$ can be written in the form of a Confluent Heun Equation (CHE) in the variable 
\beq 
z_{\rm in} = \frac{r_b-r}{r_b-r_s}\,,
\eeq 
namely
\bea\label{CHEin}
&&H''(z_{\rm in})+\left({\gamma\over z_{\rm in}}+{\delta\over z_{\rm in}-1}+\eta\right)H'(z_{\rm in})\nonumber\\
&&+{\alpha z_{\rm in}-\beta\over z_{\rm in}(z_{\rm in}-1)}H(z_{\rm in})=0\,,
\eea
with parameters 
\bea
\label{list_of_params}
\alpha &=&  2i\epsilon + 2 \epsilon \left( {\tau} - i{\kappa}\right) \,, 
\nonumber\\
\beta &=& \ell(\ell+1) + i\epsilon + {\kappa}  -{\kappa}^2 - \frac13 (\epsilon+2\tau)^2 + 2 \epsilon {\tau} 
\,,\nonumber\\
\gamma&=&  1\,,\nonumber\\
\delta&=&  1 - 2{\kappa}\,,\nonumber\\
\eta&=& 2i\epsilon\,,
\eea
where, for later convenience and comparison with Schwarzschild BHs, we have introduced the combinations 
\beq 
\epsilon=-(r_b-r_s)\omega\,, \qquad {\tau} = \omega\left(r_s+{r_b\over 2}\right)\,.
\eeq
These two relations,  linear in $r_b$ and $r_s$, imply
\beq
\omega r_b=-\frac23 (\epsilon-\tau)\,,\qquad \omega r_s=\frac13 (\epsilon+2\tau)\,,
\eeq
so that  the following identity holds
\beq
27 \kappa^2 \epsilon +(\epsilon+2\tau)^3=0\,,
\eeq
which can be used, for example, to re-express $\kappa$, Eq. \eqref{om_lam_kappa_defs}, as a function of $\epsilon$ and $\tau$.
Obviously, Eq. \eqref{tsw1pyzero} admits a second, independent solution which is of the type
$\tilde R(r)=e^{i\omega r} (r-r_s)^{+{\kappa}}\tilde H(r)$, with $\tilde H(r)$ satisfying Eq. \eqref{CHE}
with the parameters of Eqs. \eqref{list_of_params} but with $\omega \to-\omega$.

\section{Three types of solution for the main radial equation}

Let us consider three different types of solution of Eq. \eqref{tsw1}: 1) MST-type, 2) Post-Newtonian, 3) JWKB-type.

\subsection{MST-type solutions}
Let us distinguish between \lq\lq MST-type In-solutions" with regularity conditions imposed at $r_b$ and \lq\lq MST-type Up-solutions" with prescribed behaviour at infinity, respectively.

\subsubsection{MST-type In solution}

In order to adapt the MST procedure to the present 5D context, 
let us rewrite Eq. \eqref{CHEin} as follows 
\bea
\label{EqCHEinzin}
&&z_{\rm in} (1-z_{\rm in})H''(z_{\rm in})+[\gamma-(\gamma+\delta)z_{\rm in}]H'(z_{\rm in})\nonumber\\
&&\qquad +(\beta-\tilde{\beta}) H(z_{\rm in})=S_{\rm in}\,,
\eea
where the (formal) `source' term is given by
\bea
S_{\rm in}=-\eta z_{\rm in}(1-z_{\rm in})H'(z_{\rm in})+(\alpha z_{\rm in}-\tilde{\beta}) H(z_{\rm in})\,.
\eea
Here, we added to both sides $-\tilde{\beta} H(z_{\rm in})$ with $\tilde\beta$ a constant to be fixed later on. 
Comparing the LHS with the hypergeometric equation for $F(z)={}_2F_1(a,b,c,z)$, namely
\beq
z(1-z)F''(z)+[c-(1+a+b)z]F'(z)-ab F(z)=0\,,
\eeq
we identify 
\be
\label{param_defs}
\gamma=c\,,\quad \gamma+\delta= 1+a+b\,,\quad ab=\tilde{\beta}-\beta\,.
\ee
Since $\gamma=1$ ({\it i.e.} $c=1$),  the last two relations in \eqref{param_defs} yield
\be\label{discrTS}
a,b ={\delta\over2}\pm \sqrt{{\delta^2\over4}-\tilde{\beta}+\beta}\,,
\ee
In analogy with the MST treatment of the Schwarzschild case,  we require that the expressions for the coefficients $a$ and $b$ do not involve square roots, namely
\beq
{\delta^2\over4}-\tilde{\beta}+\beta=\left(\nu+\frac12 \right)^2\,,
\eeq
with $\tilde \beta$ expressed here in terms of the parameter $\nu$, that plays the role of a `renormalized' angular momentum in that $\nu =\ell + ...$.

Therefore, from Eq. \eqref{discrTS}
\beq
a=\frac{\delta+ 1}{2}+ \nu\,,\qquad
b=\frac{\delta- 1}{2}- \nu\,,
\eeq
so that Eqs. \eqref{list_of_params} then imply
\bea
a&=& 1+\nu-\kappa\,,\nonumber\\
b&=& -\nu -{\kappa}\,,\nonumber\\
c&=& 1 \,. 
\eea
Let us look for solutions of Eq. \eqref{EqCHEinzin} 
by using the ansatz
\be
H(z_{\rm in})=\sum_{n=-\infty}^{\infty}C_n F_{\nu+n}\,,
\ee
with
\beq
F_{n+\nu} = {}_2F_1(a+n,b-n,1,z_{\rm in})\,.
\eeq
Using the identities 
\be
z_{\rm in}(1-z_{\rm in})\frac{dF_{\nu+n}}{dz_{\rm in}} =A_+F_{\nu+n+1} + A_0 F_{\nu+n} +A_{-} F_{\nu+n-1}
\ee
where
\bea
A_+&=&\frac{(-\kappa +\nu +n+1) (\kappa +\nu +n) (\kappa +\nu +n+1)}{2
   (\nu +n+1) (2 \nu +2 n+1)}\,,\nn\\
   A_0&=&\frac{\kappa  (\kappa -\nu -n-1) (\kappa +\nu +n)}{2 (\nu +n)
   (\nu +n+1)}\,,\nn\\
   A_-&=&-\frac{(-\kappa +\nu +n) (-\kappa +\nu +n+1) (\kappa +\nu +n)}{2
   (\nu +n) (2 \nu +2 n+1)}\,,\nonumber\\
\eea
and
\be
z_{\rm in}F_{\nu+n} =B_+F_{\nu+n+1}+B_0 F_{\nu+n} +B_{-} F_{\nu+n-1}\,,
\ee
where
\bea
B_+&=&-\frac{(-\kappa +\nu +n+1) (\kappa +\nu +n+1)}{2 (\nu +n+1) (2
   \nu +2 n+1)}\,,\nn\\
   B_0&=&\frac{1}{2} \left(1-\frac{ \kappa ^2}{(\nu +n) (\nu
   +n+1)}\right)\,,\nn\\
   B_-&=&-\frac{(-\kappa +\nu +n) (\kappa +\nu +n)}{2 (\nu +n) (2 \nu +2
   n+1)}\,,
\eea
we find the recursion relations 
\be
\alpha_n C_{n+1}+\beta_n C_n+\gamma_n C_{n-1}=0\,,
\ee
where
\bea
\alpha_n&{=}&{-}\frac{i \epsilon  \left({\kappa} {+}n{+}\nu {+}1\right) \left({-}{\kappa}{+}n{+}\nu {+}1\right) \left(i {\tau} {+ }n{+}\nu{+}1\right)}{(n{+}\nu {+}1) (2 n{+}2 \nu {+}3)} \nonumber\\
\beta_n&=&\ell  (\ell +1)-(n+\nu ) (n+\nu +1)+\frac{\epsilon{\kappa}^2 {\tau}}{ (n+\nu ) (n+\nu +1)}\nonumber\\
&-&\frac13(4\tau^2 +  \tau\epsilon +  \epsilon^2) 
\,,\nonumber\\
\gamma_n&=&\frac{i \epsilon  \left({\kappa}+n+\nu \right) \left(-{\kappa}+n+\nu \right) \left(- i {\tau}+n+\nu
   \right)}{(n+\nu ) (2 n+2 \nu -1)}\,.
\eea
Notice the invariance under $\kappa \leftrightarrow - \kappa$ of the above coefficients.
For $n=0$ one can solve perturbatively the continuous fraction
\be
\beta _0-\frac{\alpha _{-1} \gamma _0}{\beta _{-1}-\frac{\alpha _{-2} \gamma _{-1}}{\beta _{-2}-\frac{\alpha _{-3} \gamma _{-2}}{\beta
   _{-3}-...}}}-\frac{\alpha _0 \gamma _1}{\beta _1-\frac{\alpha _1 \gamma _2}{\beta _2-\frac{\alpha _2 \gamma _3}{\beta _3-\frac{\alpha_3\gamma_4}{\beta_4-...}}}}=0\,,
\ee
leading to\footnote{This is  `different' from SW/AGT whereby $a(u)$ is expanded in powers of $\epsilon = \omega(r_s-r_b)$ only and the other parameters appear in the `masses' $\tau = \mp i m_3$ and $\kappa = m_1=\mp m_2$ and in the very definition of $u=(\ell+{1\over 2})^2 + ... $}
\be
\nu=\ell+\sum_{n=1}^\infty \nu_n \omega^n\,.
\ee 
The coefficients $\nu_n$ are  such that $\nu_{2n+1}=0$ for $n=0,1,2,...$ and, for example,
\bea
\nu_2&=&N_{20}(\ell)r_b^2 +N_{11}(\ell)r_b  r_s+N_{02}(\ell)r_s^2\nonumber\\
&\equiv &\frac{(2-3L) r_b^2+(5-6 L) r_b r_s+(11-15L) r_s^2}{2 (2 \ell -1) (2 \ell +1) (2 \ell +3)}\,,\nonumber\\
\eea
where we recall $L=\ell(\ell+1)$. Similarly,
\bea
\nu_4&=&N_{40}(\ell)r_b^4 +N_{31}(\ell)r_b^3 r_s+N_{22}(\ell)r_b^2r_s^2\nonumber\\
&+& N_{13}(\ell)r_br_s^3+N_{04}(\ell)r_s^4\,,
\eea
with the terms $N_{ab}(\ell)$ now rather involved functions of $\ell$.
The coefficients $\nu_n$, however, match with the corresponding Schwarzschild coefficients  for $s=0$ in the limit $r_b=0$.

\subsubsection{MST-type  Up solution}

In order to study the solution of Eq. \eqref{tsw1pyzero} at radial infinity let's switch the independent variable to
\be
z_{\rm up} = \tilde{z}=\omega(r-r_s)\,.
\ee
Eq. \eqref{tsw1pyzero} then becomes  
\bea
&&\tilde{z}(\tilde{z}+\epsilon)R''(\tilde{z})+(2\tilde{z}+\epsilon)R'(\tilde{z})+\left(\tilde{z}^2-{3r_s\epsilon \tilde{z}\over r_b-r_s}\right.\nonumber\\
&{+}&\left. {3r_s^2\epsilon^2\over (r_b{-}r_s)^2}{-}{r_s^3\epsilon^3\over (r_b{-}r_s)^3\tilde{z}}{-}\ell(\ell{+}1)\right)R(\tilde{z}){=}0\,,\qquad
\eea
where $\epsilon = \omega (r_s-r_b)$. Let us further rescale the radial function\footnote{The prefactor $1/\tilde{z} \sim 1/r$ is the natural one for spherical waves in $D=4$. Note that this rescaling does not introduce any singularity since $\tilde{z}=0$, i.e., $r=r_s$ is not part of the TS space-time.} with 
\be
R(\tilde{z})=\tilde{z}^{-1} f(\tilde{z})\,,
\ee
so that the new radial equation is 
\bea
\label{new_rad_eq}
&&\tilde{z}(\tilde{z}+\epsilon)f''(\tilde{z})-\epsilon f'(\tilde{z})+\Big[\tilde{z}^2-{3 r_s \epsilon \tilde{z}\over r_b-r_s}+{3r_s^2\epsilon^2\over(r_b-r_s)^2}\nonumber\\
&&-\ell(\ell+1)+{\epsilon\over \tilde{z}}\left(1-{r_s^3\epsilon^2\over (r_b-r_s)^3}\right)\Big]f(\tilde{z})=0\,.
\eea
Eq. \eqref{new_rad_eq} can be rewritten in the form
\bea
&&\tilde{z}^2 \Big[f''(\tilde{z})+f(\tilde{z})\Big]- \left({(2r_s+r_b)\epsilon \tilde{z}\over r_b-r_s}+\nu(\nu+1)\right)f(\tilde{z})\nonumber\\
&&\qquad=S_\epsilon(\tilde{z})\,,
\eea
where the `source' term turns out to be
\bea 
&&S_{\rm up}(\tilde{z})=-\epsilon \tilde{z} f''(\tilde{z})+\epsilon f'(\tilde{z})\nonumber\\
&&+\Big[-\epsilon \tilde{z}+\ell(1+\ell)-\nu(1+\nu)-{3 r_s^2\epsilon^2\over (r_b-r_s)^2}\nonumber\\
&&-{\epsilon\over \tilde{z}}\left(1-{r_s^3\epsilon^2\over (r_b-r_s)^3}\right)\Big]f(\tilde{z}) \,.
\eea
The solution of the associated homogeneous equation with outgoing wave condition reads
\be
f_0(\tilde{z})=(-2i\tilde{z})^{1+\nu}e^{i\tilde{z}}U\Big[1+\nu-i\tau,2(1+\nu),-2i\tilde{z}\Big]\,,
\ee
where $U$ is the Tricomi confluent hypergeometric function\footnote{Sometimes denoted by $\Psi(a,b;x)$.}, `decreasing' at infinity (no `resurgent' exponential $e^{-2ix}$ terms) and related to the `standard' confluent hypergeometric function\footnote{Sometimes denoted by $\Phi(a,b;x)$.} ${}_1F_1$ by
\bea
U(a,b;x)&=&{\Gamma(1-b)\over \Gamma(a-b+1)}{}_1F_1(a,b,x)\nonumber\\
&+&{\Gamma(b-1)\over \Gamma(a)}x^{1-b}{}_1F_1(a-b+1,2-b,x)\,.\nonumber\\
\eea
The inverse formula for ${}_1F_1$ in terms of $U$  
\bea
{1\over\Gamma(b)}{}_1F_1(a,b,x)&=&{e^{\mp a\pi i}\over \Gamma(b-a)}U(a,b,x)\nonumber\\
&+&{e^{\pm(b-a)\pi i}\over\Gamma(a)}e^x U(b-a,b,e^{\pm \pi i }x)\,,\nonumber\\
\eea
shows the `undesired' exponential in the second term, that is why we use Tricomi $U\approx \Psi$ rather than Kummer's ${}_1F_1 \approx \Phi$.

For later use, let us also notice that
\be
x^{-a}U(-a+i\eta,-2a,x) = x^{a+1}U(a+1+i\eta,2a+2,x)\,.
\ee
Thus
\bea
f_n(\tilde{z})&=&(-2i\tilde{z})^{1+\nu+n}e^{i\tilde{z}}\times \nonumber\\
&&U\Big[\nu+n+1-i\tau,2(\nu+n+1),-2i\tilde{z}\Big]\nonumber\\ 
&=& {{W}}_{\nu+n}(\tilde{z})\nonumber\\
&=& e^{i\tilde{z}} \{(-2i\tilde{z})^{1+\nu+n}  
{\Gamma(-2\nu-2n-1)\over \Gamma(-\nu-n-1-i\tau)}\times \nonumber\\
&& {}_1F_1(\nu+n+1-i\tau,2(\nu+n+1),-2i\tilde{z})\nonumber\\
&+& (-2i\tilde{z})^{-\nu-n}{\Gamma(2\nu+2n+1)\over \Gamma(\nu+n+1-i\tau)}\times \nonumber\\
&&{}_1F_1(-\nu-n-i\tau,-2(\nu+n),-2i\tilde{z})\}\,,
\eea
Let us choose as ansatz
\be
\label{tilde_f_def}
f(\tilde{z}; \nu) = \sum_{n=-\infty}^\infty \tilde{C}_n {{W}}_{\nu+n}(\tilde{z})\,,
\ee
with the Coulomb functions, defined by   
\bea
W_{\nu+n}(\tilde{z})&=&(-2i\tilde{z})^{1+n+\nu}e^{i\tilde{z}}\times \nonumber\\
&&U\Big[1+n+\nu-i{\tau},2(1+n+\nu),-2i\tilde{z}\Big]\,,\nonumber\\
\eea
satisfying the functional identities 
\begin{widetext}
\bea
{d W_{n +\nu}\over d\tilde{z}} &=& {i (\nu {+}n) \left(-i {\tau}{+}\nu {+}n{+}1\right)\over (\nu{+}n{+}1) (2 \nu {+}2 n{+}1)}{{W}}_{n{+}\nu{+}1}{+}
{\tau \over (n{+}\nu)(n{+}\nu{+}1)}{{W}}_{n{+}\nu}
{+}{i (\nu {+}n{+}1) \left({+}i{\tau}{+}\nu {+}n\right)\over (\nu {+}n) (2 \nu {+}2 n{+}1)}{{W}}_{n{+}\nu{-}1}\,,\nn\\
\frac{1}{\tilde{z}} W_{n+\nu} &=&-i {\left(-i {\tau}+\nu +n+1\right)\over (\nu +n+1) (2 \nu +2n+1)}{{W}}_{n+\nu+1} + {\tau\over (n+\nu)(n+\nu+1)}{{W}}_{n+\nu}+i {\left(i {\tau}+\nu +n\right)\over (\nu +n) (2 \nu +2 n+1)}{{W}}_{n+\nu-1}\,.\nonumber\\
\eea
\end{widetext}
Plugging $f(\tilde{z}; \nu)$ as given in Eq. \eqref{tilde_f_def} in the `inhomogeneous' wave equation one finds the following recursion relation for the coefficients $\tilde{C}_n$ of the Up-solution
\be
\tilde\alpha_n \tilde{C}_{n+1} + \tilde\beta_n \tilde{C}_{n} + \tilde\gamma_n \tilde{C}_{n-1}  = 0\,,
\ee
with
\bea
\label{SarifumaMedo}
\tilde\alpha_n&{=}&{-}\frac{i \epsilon  \left({\kappa} {+}n{+}\nu {+}1\right) \left({-}{\kappa}{+}n{+}\nu {+}1\right)\left(i {\tau} {+}n{+}\nu
    {+}1\right)}{(n{+}\nu {+}1) (2 n{+}2 \nu {+}3)} \nonumber\\
&& \nonumber\\ 
&=& \alpha_n \,,\nonumber\\
\tilde\beta_n&=&\ell  (\ell +1)-\nu  (\nu +1)-n (n+2 \nu +1)\nonumber\\
&-&\frac{\epsilon{\kappa}^2 \tau}{2 (n+\nu ) (n+\nu +1)}-\frac{\omega^2 \left(r_b r_s+r_b^2+4
   r_s^2\right)}{2}\nonumber\\
&=&\beta_n\,,\nonumber\\
\tilde\gamma_n&=&\frac{i \epsilon  \left({\kappa}+n+\nu \right) \left(-{\kappa}+n+\nu \right) \left(- i {\tau}+n+\nu
   \right)}{(n+\nu ) (2 n+2 \nu -1)}\nonumber\\
&=& \gamma_n\,.
\eea
As indicated, these exactly coincides with the coefficients in the recursion relation for the coefficients $C_n$ of the In-solution, i.e., 
\be
\tilde{C}_n = {C}_n \,,
\ee
up to an overall $n$-independent factor. 
This is slightly different but not inconsistent with what MST found in the BH context. There, in order to connect the In and Up solutions, one had to properly rescale the expansion coefficients, see {\it e.g.} Eq. (120) in Ref. \cite{Sasaki:2003xr}. 
We check that this is indeed necessary in the BH case.
The integrability condition on $\nu$ is the same, so there is only one `renormalized' angular momentum, later on related to the quantum Seiberg-Witten period $a$, for the CHE governing the dynamics of scalar perturbations of TSs.
\subsubsection{Estimating the coefficients}

Setting 
\be 
{\cal R}_n = {C_n\over C_{n-1}}\,, \qquad {\cal L}_n = {C_n\over C_{n+1}}\,,
\ee
the recursion relations yield
\be 
{\cal R}_n = - {\gamma_n\over \beta_n + \alpha_n {\cal R}_{n+1}} \quad , \quad {\cal L}_n = - {\alpha_n\over \beta_n + \gamma_n {\cal L}_{n-1}}\,,
\ee
with the integrability condition
\be 
{\cal R}_n(\nu) {\cal L}_{n-1}(\nu) = 1\,.
\ee
that is an `eigenvalue' equation for $\nu$ and allows to express it in terms of $\ell$, $\omega$ and the length parameters $r_b$ and $r_s$.

Noticing that for large $|n|$
\be
\alpha_n \approx -{i\over 2} \epsilon n\,, \quad \beta_n \approx n^2\,, \quad \gamma_n \approx +{i\over 2} \epsilon n\,,
\ee
one easily finds 
\be
\lim_{n\rightarrow +\infty} n {C_n\over C_{n-1}} = - {i\over 2} \epsilon \,, \quad 
\lim_{n\rightarrow -\infty} n {C_n\over C_{n+1}} = + {i\over 2} \epsilon\,.
\ee
For $n\ge 1$ one can use the recurrence for ${\cal R}_n$ and find $C_n\approx (-i\epsilon)^n/2^n n!$. 
For $n<0$ one should take into account the vanishing to order zero in $\epsilon$ of $\alpha_n$ for $n=-\ell-1$ (and $n=-\ell-{3\over 2}$ for half-integer $\ell$) and of  $\beta_n$ for $n=-2\ell-1$ when $\nu\approx \ell$. As a result, while generically ${\cal L}_n \approx \epsilon$, ${\cal L}_{-\ell-1} \sim {\cal O}(1)$ and ${\cal L}_{-2\ell-1} \sim 1/\epsilon$ and then
\beq
C_n  \sim \left\{\begin{array}{ll}
 \epsilon^{|n|}  & {\rm for}\quad -1\ge n \ge -\ell\,,\cr
  \epsilon^{\ell} & {\rm for}\quad n = {-\ell-1}\,,\cr
  \epsilon^{|n|-1} & {\rm for}\quad -\ell-2\ge n \ge -2\ell\,,\cr
 \epsilon^{2\ell-2}& {\rm for}\quad n = {-2\ell-1}\,,\cr
 \epsilon^{|n|-3} & {\rm for}\quad {-2\ell-2}\ge n\,.
\end{array}
\right.
\eeq
Thanks to the highly suppressed behavior of the coefficients, the series of (confluent) hypergeometric functions converge very well.

\subsection{Post-Newtonian solutions}

Limiting our considerations to the case $P_y=0$, for simplicity, let us now discuss the Post-Newtonian (PN) expansion of the solutions. To this end we
add a PN weight $\eta_v=1/c$ to the parameters entering Eq. \eqref{tsw1} 
$r_s\to \eta_v^2 r_s$, $r_b\to \eta_v^2 r_b$, $\omega\to \omega \eta_v$.
One can justify this rescaling via the following argument. For a massive probe with velocity $v$ one has $v^2\sim GM/r$ so for fixed $r=r_0>>GM/c^2$ (as needed later) $GM/c^2\sim r_s \sim r_b \sim r_0 v^2/c^2$, so $r_{s,b} \sim \eta_v^2$. On the other end $\omega/c \sim v/c r_0 \sim \eta_v$.

In other terms, using $\omega r = v$ and $v^2 \sim GM/r$ as well as $GM/c^2\sim r_b \sim r_s$ (since $1<r_b/r_s <2$), one finds
\bea
z_{\rm up}&=&\omega (r-r_s)/c \sim v/c - (v/c)^3\,,\nonumber\\
\epsilon, \tau, \kappa &\sim& \omega GM/r c^2= GM\omega^2/ \omega r c^2  \sim v^3/c^3
\eea
While $\epsilon, \tau, \kappa\sim v^3/c^3$ all scale with the same power of $v/c = \eta_v$, $z_{\rm up}$ contains two different terms scaling with different powers of 
$v/c = \eta_v$. With this in mind one has
\bea
\label{tsw2}
&&\frac{d^2R}{dr^2}+\left(\frac{1}{r-r_b\eta_v^2}+\frac{1}{r-r_s\eta_v^2}\right)\frac{dR}{dr}\nonumber\\
&+&\frac{r^3}{(r-r_s\eta_v^2)(r-r_b\eta_v^2)}\left(\frac{\omega^2\eta_v^2}{r-r_s\eta_v^2}- \frac{\ell(\ell+1)}{r^3} \right) R(r)=0\,,\nonumber\\
\eea
Expanding in powers of $\eta_v$ the equation and the solution
\beq
R_{\rm PN}(r)=\sum_{k=0}^\infty R_k(r)\eta_v^k\,,
\eeq 
we find two independent solutions, one regular at the origin ($R_{\rm PN, in}(r)=R_{\rm PN, in}(r; r_b,r_s,\ell,\omega)$) and the other ($R_{\rm PN, up}(r)=R_{\rm PN, up}(r; r_b,r_s,\ell,\omega)$) regular at infinity, namely
\bea
&&R_{\rm PN, in}(r; r_b,r_s,\ell,\omega)=r^\ell \nonumber\\
&-& \left(\frac{(r_b+r_s)\ell}{2}+\frac{r^3\omega^2}{2(2\ell+3)}\right) \eta_v^2 r^{\ell-1}\nonumber\\
&+& {\cal O}(\eta_v^4)\\
&&R_{\rm PN, up}(r; r_b,r_s,\ell,\omega)=R_{\rm PN, in}(r; r_b,r_s,-\ell-1,\omega)\nonumber\\
&=& r^{-\ell-1}+\left(\frac{(r_b+r_s)(\ell+1)}{2}-\frac{r^3\omega^2}{2(2\ell+1)}\right) \eta_v^2 r^{-\ell-2}\nonumber\\
&+&{\cal O}(\eta_v^4)\,. \nonumber
\eea
One can form their rescaled (and thus constant) Wronskian
\bea
&&W_{\ell m; \omega, r_b,r_s}= r^2f_s(r)f_b(r){\widetilde{W}}_{\rm in-up}(r)\nonumber\\ 
&&= -(1+2\ell)\left(1+W^{\eta_v^6}\omega^2 \eta_v^6 \right)+{\cal O}(\eta_v^8)\,,
\eea
where the `true' (non-constant but `nowhere' vanishing) Wronskian reads
\bea
{\widetilde{W}}_{\rm in-up}(r)&=&R_{\rm PN, in}(r)R_{\rm PN, up}'(r) -R_{\rm PN, up}(r)R_{\rm PN, in}'(r) \nn \\
\eea
and 
\bea
W^{\eta_v^6}&=& W_{20}^{\eta_v^6}r_b^2+W_{11}^{\eta_v^6}r_br_s+W_{02}^{\eta_v^6}r_s^2\,,
\eea
with
\bea
W_{20}^{\eta_v^6}&=& -\frac{(60 \ell^4+120 \ell^3-33 \ell^2-93 \ell+38)}{8 (2 \ell+3)^2 (-1+2 \ell)^2}\,,  \nonumber\\
W_{11}^{\eta_v^6}&=& -\frac{(84 \ell^4+168 \ell^3-43 \ell^2-127 \ell+46 ) }{ 4(2 \ell+3)^2 (-1+2 \ell)^2}\,,\nonumber\\
W_{02}^{\eta_v^6}&=& -\frac{(124 \ell^4+248 \ell^3-65 \ell^2-189 \ell+74)}{ 8(2 \ell+3)^2 (-1+2 \ell)^2} \,.
\eea
These PN solutions (for which we have displayed the first two terms) form a basis of solutions, with the nice property that the Up solutions can be simply obtained by the corresponding in ones by replacing $\ell\to {-}\ell{-}1$. However, we need solutions regular at the `cap' $r_b$ (or $r_s$ for BHs) and infinity, and these are of the type Mano-Suzuki-Takasugi (MST) found originally for the BH case, as already discussed.

\subsection{JWKB solutions}

Another often used approach is the semi-classical JWKB approximation for which   various expressions are available, see {\it e.g.} Sec. IV in \cite{Bini:2015mza}.
For example, in Eq.  \eqref{tsw1pyzero} one can switch to the new, tortoise-like, variable\footnote{This choice is not unique since it follows by solving a second order ODE.} 
\beq
r_*=(r_b-r_s)\left[\ln\left(\frac{ r-r_b }{ r-r_s }\right)+1\right]\,,
\eeq
with 
\beq
\frac{dr_*}{dr}=\frac{(r_b-r_s)^2}{(r-r_b) (r-r_s)}\,,
\eeq
and inverse
\beq
r=\frac{r_s{\zeta}-r_b}{{\zeta}-1}\,,\qquad {\zeta}=e^{1+\frac{r_*}{r_{b}-r_s}}\,,
\eeq
such that when $r\to r_b$ one has $r_*\to -\infty$ while when $r\to \infty$  one has $r_*\to r_b-r_s$.
Eq.  \eqref{tsw1pyzero} then becomes
\beq
\frac{d^2}{dr_*^2}R(r_*)-L\frac{r-r_b}{(r_b-r_s)^4}\left[ (r-r_s)-\frac{\omega^2}{L^2} r^3\right]R(r_*)=0\,.
\eeq
Assuming $\omega=m\Omega$ and defining $w=m/L$ (assumed to be constant when $L\to \infty$) we find
\beq
\frac{d^2}{dr_*^2}R(r_*)-L\frac{r-r_b}{(r_b-r_s)^4}\left[ (r-r_s)-w^2\Omega^2 r^3\right]R(r_*)=0\,.
\eeq
Writing
\beq
L=\frac{1}{\hbar^2}-\frac14\,,
\eeq
(with $\hbar$ just a useful symbol here and not the Planck constant)  and expanding in $\hbar \to 0$
\beq
\label{WKB_pronta}
\frac{d^2}{dr_*^2}R(r_*)-\frac{Q(r_*)}{\hbar^2} R(r_*)=0\,.
\eeq
with
\beq
Q(r_*)=\frac{r-r_b}{(r_b-r_s)^4}\left[ (r-r_s)-w^2\Omega^2 r^3\right]\bigg|_{r=r(r_*)}\,.
\eeq
The WJKB theory allows one to write the solutions $R(r_*)$ in the form
\beq
R(r_*)=e^{\frac{S_0}{\hbar}+S_1+O(\hbar)}\,,
\eeq
where
\beq
\frac{S_0}{\hbar}=\pm \int p(r_*)d r_*\,,\qquad S_1=-\frac12 \ln p(r_*)
\eeq
with
\beq
p(r_*)=\sqrt{Q(r_*)}\,.
\eeq
Correspondingly, the two independent solutions of Eq. \eqref{WKB_pronta} are given by
\beq
R_\pm (r_*)=C_\pm \frac{e^{\pm \int^{r_*} p(r_*)d r_*}}{\sqrt{p(r_*)}}\,,
\eeq
and can be evaluated, for example, in PN sense. 
The choice $C_\pm=\frac{1}{\sqrt{2}}$
would imply that the Wronskian of
these solutions is $1$ 
\beq
W=R_-(r_*)\frac{d}{dr_*}R_+(r_*)-R_+(r_*)\frac{d}{dr_*}R_-(r_*)=1\,.
\eeq
Finally, let us note that
\beq
G(r_*)=\frac{R_+(r_*) R_-(r_*)}{W}=\frac{C_+C_-}{p(r_*)}=\frac{1}{2\sqrt{Q(r_*)}}\,.
\eeq

For the sake of comparison, let us notice that in \cite{Heidmann:2023ojf} a different JWKB approximation is adopted in which the wave function is $R(r)=\psi(r)/(r-r_s)$ and the independent tortoise-like variable is $\tilde{r}_*=r-r_b+(r_b-r_s)\ln [(r-r_b)/r_o]$, where $r_o$ is an arbitrary constant. Independently of $r_o$, $r\rightarrow \infty$ corresponds to $\tilde{r}_*\rightarrow \infty$, while the `cap' at $r\rightarrow r_b$ corresponds to $\tilde{r}_*\rightarrow - \infty$.

\section{Scalar field sourced by a scalar charge}

As a proxy for GW emission from TSs, we study massless  scalar wave emission 
by a low-mass ($\mu\ll M_{TS}$) non-spinning neutral ($P_y=0$) probe orbiting around a TS on the ISCO, i.e. a  stable time-like circular `critical' geodesic outside the light-ring.

A similar problem has been recently address by \cite{Barack:2023oqp} in the context of a toy model with a massless scalar $\psi$ that interacts with a low-mass ($m_1$) scalar $\phi_1$ via a trilinear coupling ${\cal L}_3 = 2\sqrt{\pi} m_1 Q \psi \phi_1^2$ in presence of a large mass ($m_2$) scalar $\phi_2$, mimicking the background of a Schwarzschild BH.

Thanks to spherical symmetry of TSs the geodesic may be taken to lay on the equatorial plane $\theta = \pi/2$ at a radius $r=r_0$ that depends on the angular momentum $J$ and energy $E$ of the probe. 

Let us recall here, for the reader's convenience, that angular velocity $\Omega=\frac{d\phi}{dt}$ and relativistic factor $\Gamma$ are given in Eqs. \eqref{Omega_circ} and \eqref{E_circ}, namely,
\beq
\Omega = \sqrt{\frac{r_s}{2r_0^3}} \,,\qquad \Gamma =\frac{1}{\sqrt{1-\frac{3r_s}{2r_0}}}\,,
\eeq
Then, the inhomogeneous wave equation to solve is
\be
\Box \phi=-4\pi \rho\,,
\ee
where the scalar charge density $\rho$ has support on a time-like worldline with parametric equations $z^\alpha=z^\alpha(\tau)$. In the present case, we consider the orbit to be a circular geodesic at radius $r=r_0>r_b$ as discussed in Section \ref{TSmetric}, so that it reads
\beq
\label{dens}
\rho = Q_{\rm S}\int_{-\infty}^\infty {1\over \sqrt{-g}}\delta^{(5)}(x^\alpha-z^\alpha(\tau))d\tau\,,
\eeq
where $Q_{\rm S}$ is the scalar coupling constant\footnote{{\it viz.} $Q_{\rm S}= 2\sqrt{\pi} m_1 Q$ in the toy model of \cite{Barack:2023oqp}.} 
\bea
\delta^{(5)}(x^\alpha-z^\alpha(\tau))&=&\delta(t-\Gamma \tau)\delta(r-r_0)\times \nonumber\\
&& \delta(\theta-\pi/2)\delta(\varphi-\Omega  t)\delta(y)\,.
\eea
After the integration over $\tau$
\bea
\rho&=&\frac{Q_{\rm S}}{\Gamma r_0^2}\delta(r-r_0)\delta(\theta-\pi/2)\delta(\varphi-\Omega t)\delta(y)\,,
\eea
where  we have used $\sqrt{-g}=r^2\sin\theta$. 
Furthermore, Eq. \eqref{dens} can be rewritten as 
\bea
\rho&=&{Q_{\rm S}\over \Gamma r_0^2}\delta(r-r_0)\delta(y)\times \nonumber\\ 
&&\sum_{\ell=0}^\infty\sum_{m=-\ell}^\ell Y_{\ell,m}(\theta,0)Y_{\ell,m}^*({\pi\over2},0)e^{-i(\omega t-m \varphi)}\,,
\eea
and $\delta(y)$  is re-expressed as a sum over KK momenta $P_y= n/R_y = p$
\be
\delta(y)=\sum_{n}{e^{i {n\over R_y}y}\over 2\pi R_y}\,.
\ee
Notice that the $\rho^{\rm 5D}$ has not the same dimensions than the  $\rho^{\rm 4D}$. As a consequence, the scalar field $\phi^{\rm 5D}\sim \frac{Q_S}{L^2}$ whereas
 $\phi^{\rm 4D}\sim \frac{Q_S}{L}$, with $L$ a generic length scale. This feature should be taken into account when comparing, for example, with 4D black hole perturbations. 

Expanding also $\Phi$ in spherical harmonics the equation to be solved (for $P_y=0$) is then
\bea
&&R''(r)-{r_s+r_b-2r\over (r-r_b)(r-r_s)}R'(r)\nonumber\\
&&+{\omega^2 r^3-\ell(\ell+1)(r-r_s)\over (r-r_s)^2(r-r_b)}R(r)\nonumber\\
&&=-{2Q_{\rm S}\over \Gamma} {Y^*_{\ell,m}(\pi/2,0)\over (r_0-r_b)(r_0-r_s)}\delta(r-r_0)\,,
\eea
namely the previous Eq.\eqref{tsw1}, with a Dirac-delta localized source term at $r=r_0$,
\bea
S&=&-{2Q_{\rm S}\over \Gamma}\delta(r-r_0) {Y^*_{\ell,m}(\pi/2,0)\over (r_0-r_b)(r_0-r_s)}\nonumber\\
&\equiv & S_{\ell m\omega}\delta(r - r_0)\,.
\eea
Let us introduce 
the Green function $G_{\ell m\omega}(r,r')$ which satisfies the equation 
\bea
&&\partial_{r}^2G_{\ell m\omega}(r,r') -{r_s+r_b-2r\over (r-r_b)(r-r_s)}\partial_r G_{\ell m\omega}(r,r')\nonumber\\
&&+{\omega^2 r^3-\ell(\ell+1)(r-r_s)\over (r-r_s)^2(r-r_b)}G_{\ell m\omega}(r,r')\nonumber\\
&&=r'^2f_s(r')f_b(r')\delta(r-r')
\eea
and reads
\begin{eqnarray}
G_{\ell m\omega}(r,r')&=&\frac{1}{W_{\ell m\omega}}\left[R_{\rm in}^{\ell m\omega}(r)R_{\rm up}^{\ell m\omega}(r')H(r'-r) \right.\nonumber\\
&+&\left. R_{\rm in}^{\ell m\omega}(r')R_{\rm up}^{\ell m\omega}(r)H(r-r') \right]\,,
\end{eqnarray}
where $H(x)$ denotes the Heaviside step function, $R_{\rm in}^{\ell m\omega}(r)$ and $R_{\rm up}^{\ell m\omega}(r)$ are two independent homogeneous solutions of the radial wave equation having the correct behavior at the `cap' and at infinity, respectively, and  
\bea
W_{\ell m\omega}&=&r^2 f_s(r)f_b(r)\left[R_{\rm in}^{\ell m\omega}(r)R'{}_{\rm up}^{\ell m\omega}(r)\right.\nonumber\\
&-&\left.R'{}_{\rm in}^{\ell m\omega}(r)R_{\rm up}^{\ell m\omega}(r)\right]
\eea
is the associated (constant) Wronskian.
We find then
\begin{eqnarray}
R_{\ell m\omega}(r) &=& \int G_{\ell m\omega}(r,r')r'^2 f_s(r')f_b(r')S_{\ell m\omega}\delta(r' - r_0)dr'\nonumber\\
&=& G_{\ell m\omega}(r,r_0)r_0^2 f_s(r_0)f_b(r_0)S_{\ell m\omega}\,,
\end{eqnarray}
and
\begin{eqnarray}
\label{psi0_coords}
\phi(t,r,\theta,\varphi)&=& r_0^2 f_s(r_0)f_b(r_0)\sum_{\ell m}Y_{\ell m}(\theta,0)\times \nonumber\\
&&\int\frac{d\omega}{2\pi} G_{\ell m\omega}(r,r_0)S_{\ell m\omega} e^{-i\omega t+im\varphi}\,.\qquad
\end{eqnarray}
Evaluating  $\phi(t,r,\theta,\varphi)\to \phi(t,r_0,\frac{\pi}{2},\Omega t)$ along the particle world-line 
leads to
\bea
\label{psi0def}
\phi(t,r_0,\frac{\pi}{2},\Omega t)&=&- r_0^2 f_s(r_0)f_b(r_0){2 Q_{\rm S} \over \Gamma}\times \nonumber\\
&&\sum_{\ell m}|Y_{\ell m}(\frac{\pi}{2},0)|^2{1\over (r_0-r_b)(r_0-r_s)}\nonumber\\
&&\int\frac{d\omega}{2\pi} G_{\ell m\omega}(r_0,r_0)  e^{-i\omega t+im\Omega t}\nonumber\\
&=& - {2 Q_{\rm S} \over \Gamma}\sum_{\ell m}|Y_{\ell m}(\frac{\pi}{2},0)|^2 \times \nonumber\\ 
&&\int\frac{d\omega}{2\pi} G_{\ell m\omega}(r_0,r_0)  e^{-i(\omega -m\Omega) t}\,.
\eea
Averaging in $t$ finally gives
\beq
\langle \phi \rangle =\int dt \phi(t,r_0,\frac{\pi}{2},\Omega t)\,,
\eeq
and thus 
\bea
\langle \phi \rangle&=&- {2Q_{\rm S}\over \Gamma}\times \nonumber\\
&&\sum_{\ell m}|Y_{\ell m}(\frac{\pi}{2},0)|^2  G_{\ell m\omega}(r_0,r_0)\bigg|_{\omega=m\Omega}\,.
\eea
The above expression   actually requires taking the limit $r\to r_0^\pm$ properly, and must be suitably regularized in order to remove its
singular part, because the field has a divergent behavior there.
More in detail
\bea
G_{\ell m\omega}(r_0,r_0)\bigg|_{\omega=m\Omega}&=&G_0(\ell,r_0)+G_2(\ell,r_0) m^2\nonumber\\
&+& G_4(\ell,r_0) m^4+G_6(\ell,r_0) m^6+\cdots\nonumber\\
&=& \sum_{n=0}^\infty G_{2n}(\ell,r_0)m^{2n}\,,
\eea
where we used the fact that $\Omega$ is itself a function of $r_0$.
For example, we find
\bea
G_0(\ell,r_0)&=& -\frac{1}{r_0(2\ell+1)} 
-\frac{(r_b+r_s)}{2r_0^2(2\ell+1)}\eta_v^2\nonumber\\
&-&\frac{((2 r_s r_b+3 r_s^2+3 r_b^2) L-2 r_s^2-2 r_b^2-2 r_s r_b)}{2 r_0^3 (3+2\ell) (2\ell+1) (2 \ell-1)}\eta_v^4\nonumber\\
&+&O(\eta_v^6)\,,\nonumber\\
G_2(\ell,r_0)&=&   -\frac{ r_s}{r_0^2 (2\ell+1) (2\ell-1) (3+2\ell)}\eta_v^2\nonumber\\
&-&\frac{(2 (3 r_s+r_b) L-6 r_s-3 r_b) r_s}{4r_0^3 L (2 \ell+1) (2 \ell-1) (3+2 \ell) }  \eta_v^4\nonumber\\
&+& O(\eta_v^6)\,, \nonumber\\
G_4(\ell,r_0)&=& -\frac{3 r_s^2}{2 r_0^3 (5+2\ell) (4 \ell^2-1) (4 \ell^2-9)} \eta_v^4 \nonumber\\
&+&O(\eta_v^6)\,, 
\eea
with $L=\ell (\ell+1)$.
Let us denote
\beq
{\cal M}_k(\ell)=\sum_{m=-\ell}^{+\ell} |Y_{\ell m}(\frac{\pi}{2},0)|^2 m^{k}\,,
\eeq
that are non-zero only for even values of $k$ and admit well-known expressions such as 
\bea
{\cal M}_0(\ell)&=&\frac{2\ell +1}{4\pi} \,,\nonumber\\
{\cal M}_2(\ell)&=& \frac{ (2 \ell + 1)\ell (\ell + 1) }{8\pi}\,,\nonumber\\
{\cal M}_4(\ell)&=& \frac{\ell(\ell+1) (2\ell+1) (3\ell^2+3\ell-2)}{32 \pi}\,,\nonumber\\
{\cal M}_6(\ell)&=&  \frac{\ell(\ell+1) (2\ell+1) (5\ell^4+10\ell^3-5\ell^2-10\ell+8) }{64\pi} \,, \nonumber\\
\eea
In terms of ${\cal M}_k(\ell)$, we find
\bea
\label{aver_phi}
\langle \phi \rangle&=&\sum_\ell \langle \phi \rangle_\ell\nonumber\\
&=& - {2Q_{\rm S}\over \Gamma}\sum_{\ell}[G_0(\ell,r_0){\cal M}_0(\ell)+G_2(\ell,r_0) {\cal M}_2(\ell)\nonumber\\
&+& G_4(\ell,r_0){\cal M}_4(\ell)+G_6(\ell,r_0) {\cal M}_6(\ell)+\cdots] \nonumber\\
&=& - {2Q_{\rm S}\over \Gamma}\sum_{\ell=0}^\infty G_{2\ell}(\ell, r_0) {\mathcal M}_{2\ell}(\ell)
\,.
\eea
This sum (over $\ell$) in general diverges since it incorporates also the singular part of the field.
In the limit of large $\ell$ the terms in square brackets in Eq. \eqref{aver_phi} are of the type
\bea
&& C_2(r_0) \ell^2 +C_1 (r_0)\ell +C_0(r_0)+\frac{C_{-1}(r_0)}{\ell}\nonumber\\
&& + \frac{C_{-2}(r_0)}{\ell^2}+\frac{C_{-3}(r_0)}{\ell^3}+\ldots  \,,
\eea
so that in order to obtain a finite result one has to subtract the singular field, a process known as \lq\lq mode sum regularization".
Here the subtraction term turns  out to be $B=C_0(r_0)$ 
\bea
B&=& \frac{Q_{\rm S}}{2r_0\pi}+
\frac{Q_{\rm S}}{ \pi r_0^2} \left( -\frac{1}{16}r_s+\frac{1}{4}r_b \right)\eta_v^2\nonumber\\
&+& \frac{Q_{\rm S}}{ \pi r_0^3}  \left(-\frac{1}{32} r_s r_b+\frac{3}{16} r_b^2-\frac{39}{512} r_s^2\right)\eta_v^4 \nonumber\\
&+&  \frac{Q_{\rm S}}{ \pi r_0^4}  \left(  \frac{5}{32} r_b^3-\frac{385}{4096} r_s^3-\frac{3}{128} r_s r_b^2-\frac{39}{1024} r_s^2 r_b\right)\eta_v^6\nonumber\\
&+& \frac{Q_{\rm S}}{ \pi r_0^5}  \left( -\frac{385}{8192} r_s^3 r_b+\frac{35}{256} r_b^4-\frac{5}{256} r_s r_b^3\right.\nonumber\\
&-&\left.\frac{117}{4096} r_s^2 r_b^2-\frac{61559}{524288} r_s^4\right)\eta_v^8\nonumber\\
&+& O(\eta_v^{10})\,.
\eea
This quantity can be computed directly by using the JWKB approach\footnote{
The JWKB approach  implies replacing summation over $m$ with an integral over $W=m/L\in[-1,1]$ of the form $\int_{-1}^1{dW}/{2\sqrt{1-W^2}\sqrt{(r-r_b)(r-r_s-W^2\Omega^2 r^3)}}$, so that $|Y(\frac{\pi}{2},0)|^2\to 1/\pi^2 \sqrt{1-W^2}$ and $G(r_0)=R_{\rm in}(r_0)R_{\rm up}(r_0)/W\to 1/{\sqrt{(r-r_b)(r-r_s-W^2\Omega^2 r^3)}}$.
} and the result (before PN expansion) turns out to be
\beq
B=\frac{Q_S}{\pi^2 \Gamma \Omega r_0^{3/2} \sqrt{r_0-r_b}} {\rm EllipticF}\left(\xi,\frac{1}{\xi}\right)\,.
\eeq
where
\beq
\xi=\Omega\sqrt{\frac{ r_0^3}{ r_0-r_s }}=\sqrt{\frac{r_s}{ 2(r_0-r_s) }}\,.
\eeq
The final expression (after subtracting the singular field)  in the limit of large $\ell$  is  of the type 
\bea
\left[\frac{C_{-2}(r_0)}{\ell^2}+\frac{C_{-3}(r_0)}{\ell^3}\ldots\right] \,,
\eea
and converges.
When one is not looking at the large $\ell$ limit, the terms in Eq. \eqref{aver_phi} admit a PN expansion.
Explicitly we find
\bea
T_\ell&=&\langle \phi \rangle_\ell-B\nonumber\\
&=&T_\ell^{\eta_v^2}\eta_v^2 +T_\ell^{\eta_v^4}\eta_v^4+T_\ell^{\eta_v^6}\eta_v^6+O(\eta_v^8) \nonumber\\
&=& \frac{3 Q_{\rm S} r_s}{ 16 r_0^2\pi (3+2 \ell ) (2 \ell-1) }\eta_v^2\nonumber\\
&+& O(\eta_v^4)\,.
\eea
It happens that for the first two terms (explicitly given above) one can perform the summation over $\ell$, but this gives trivially zero because the summations are telescopic,
\beq
\sum_{\ell=0}^\infty T_\ell^{\eta_v^2}=0\,,\qquad
\sum_{\ell=0}^\infty T_\ell^{\eta_v^4}=0\,.
\eeq
The next term, $O(\eta_v^6)$, involving a denominator $1/(\ell-1)$, can only be summed from $\ell=2$ up to infinity
\bea
\sum_{\ell=2}^\infty T_\ell^{\eta_v^6}&=&
\frac{Q_{\rm S}}{r_0^4}\left[\left(-\frac{1}{512}r_s r_b^2-\frac{7}{512}r_s^3-\frac{1}{128}r_s^2 r_b\right) \pi\right.\nonumber\\
&+&\left(\frac{1}{80}r_b^3+\frac{13}{14400}r_s r_b^2+\frac{21799}{806400}r_s^2 r_b\right. \nonumber\\
&+&\left.\left.\frac{126253}{1382400}r_s^3\right)\frac{1}{\pi}\right]\eta_v^6\,.
\eea
The next term, $O(\eta_v^8)$, involving a denominator $1/(\ell-2)$, can only be summed from $\ell=3$ up to infinity
\bea
\sum_{\ell=3}^\infty T_\ell^{\eta_v^8}&=&\frac{Q_{\rm S}}{r_0^5}\left[\left(-\frac{59}{16384} r_s^2 r_b^2-\frac{49}{8192} r_s^3 r_b+\frac{29}{16384} r_s^4
\right.\right.\nonumber\\
&-&\left.\frac{1}{1024} r_s r_b^3\right)\pi +\left(-\frac{4938262487443}{86790832128000} r_s^4\right.\nonumber\\
&+&\frac{13}{1280} r_b^4+\frac{16280977}{526848000} r_s^2 r_b^2 \nonumber\\
&-&\left. \left. \frac{94452697}{104315904000} r_s^3 r_b-\frac{7927}{2822400} r_s r_b^3\right)\frac{1}{\pi} \right]\eta_v^8\,.\nonumber\\
\eea
Therefore, if one is interested in a result which is accurate up to $O(\eta_v^6)$ included,  the two MST type contributions for $\ell=0$ and $\ell=1$
(up to $O(\eta_v^6)$ should be added too
\beq
\sum_{\ell=0}^\infty (\langle \phi \rangle_\ell-B)
= T_0^{\eta_v^6 \rm (MST)}+T_1^{\eta_v^6 \rm (MST)}+\sum_{\ell=2}^\infty T_\ell^{\eta_v^6}\,.
\eeq
We find
\bea
T_0^{\eta_v^6 \rm (MST)}&=&\frac{Q_{\rm S}}{r_0^2\pi}\left[ -\frac{r_s}{16}  \eta_v^2\right.\nonumber\\
&+&\left(-\frac{1}{48} r_b^2 + \frac{1}{96} r_b r_s - \frac{131}{1536} r_s^2\right)\frac{1}{r_0}\eta_v^4\nonumber\\
&+&\left(-\frac{1}{32} r_b^3 + \frac{3}{128} r_b^2 r_s - \frac{33}{1024} r_b r_s^2\right.\nonumber\\
& -&\left.\left. \frac{335}{4096} r_s^3\right)\frac{1}{r_0^2}\eta_v^6 \right]\,,\nonumber\\
T_1^{\eta_v^6 \rm (MST)}&=&
 \frac{3 Q_{\rm S} r_s}{80\pi r_0^2} \eta_v^2 \nonumber\\
&+& \frac{Q_{\rm S}}{80\pi r_0^3 } \left(r_b^2 - \frac72 r_b r_s - \frac{1315}{224} r_s^2\right)\eta_v^4 \nonumber\\
&-& \frac{Q_{\rm S}}{24\pi r_0^4}\left[\left(2 {\mathcal L}- \frac{309619}{115200}\right) r_s^3\right.\nonumber\\
&+&\left. r_b \left({\mathcal L} - \frac{35459}{22400}\right) r_s^2 + \frac{ 701 r_b^2 r_s}{1200} - \frac{9 r_b^3}{20}\right]\eta_v^6\,,\nonumber\\ 
\eea
having defined
\beq
{\mathcal L}=\ln \left(\frac{\sqrt{2}e^{\gamma_E}}{\sqrt{r_0r_s}} \right)\,.
\eeq

Similarly, if one is interested in a result which is accurate up to $O(\eta_v^8)$ included,  the three MST type contributions for $\ell=0$, $\ell=1$ and for $\ell=2$ (up to $O(\eta_v^8)$) should be added too
\beq
\sum_{\ell=0}^\infty (\langle \phi \rangle_\ell-B)
= T_0^{\eta_v^8 \rm (MST)}+T_1^{\eta_v^8 \rm (MST)}+T_2^{\eta_v^8 \rm (MST)}+\sum_{\ell=3}^\infty T_\ell^{\eta_v^8}\,,
\eeq
etc.

We can rewrite the above relation by introducing the dimensionless quantities
\beq
y=\frac12 \frac{r_s}{r_0}\,,\qquad \alpha=\frac{r_b}{r_s}\,.
\eeq
that could be used in a comparison with the corresponding results for Schwarschild BHs. 
Here we have imposed for the perturbing scalar waves regularity conditions at $r=r_b>r_s$ and purely outgoing at infinity. The limit $r_b\to 0$ reproduces the Schwarschild metric, but then also imposes regularity conditions at $r=0$, whereas in the Schwarschild case purely ingoing conditions are to be imposed at $r=r_s$. This implies that the Schwarzschild limit of the scalar waves cannot be (fully) re-obtained within our framework. It is  however useful to recall here the Schwarzschild result (in units of $Q_{\rm S}$)\\
\beq
\psi_0^{\rm reg}\sim -y^3+\left(\frac{35}{8}-\frac{7}{32}\pi^2-\frac23 \ln \left(4y e^{2\gamma_E} \right) \right)y^4+O(y^5)\,.
\eeq
with $y=\frac{M}{r_0}=\frac12 \frac{r_s}{r_0}$\,.

\section{Scalar field modifications to the background geodesics}

Quinn \cite{Quinn:2000wa}  has studied  the interaction of a scalar field $\phi$
with a particle of constant bare mass $\mu_0$ and constant
scalar charge $Q_{\rm S}$ and has shown that the equation of motion is
\beq
\label{eq_sc_mu}
U^\nu \nabla_\nu (\mu U_\lambda)=\nabla_\lambda \phi\,,
\eeq
where $\mu=\mu_0-Q_{\rm S}\phi$, that is the change in $\mu$ is of the first order in $Q_{\rm S}$,
\beq
\frac{\Delta \mu}{\mu_0}=\frac{ \mu-\mu_0}{\mu_0}=-\frac{Q_{\rm S}}{\mu_0}\phi\,.
\eeq
Projecting Eq. \eqref{eq_sc_mu} orthogonally to $U$ leads to the following acceleration-equals-force equation
\beq
a(U)^\lambda=U^\mu \nabla_\mu U^\lambda =\frac{Q_{\rm S}}{\mu}P(U)^{\lambda\mu}\nabla_\mu \phi=\frac{1}{\mu}F_{\rm scal}(U)^\lambda\,,
\eeq
with
\beq
F_{\rm scal}(U)^\lambda=Q_{\rm S} P(U)^{\lambda\mu}\nabla_\mu \phi\,,
\eeq
which modifies the worldline of the particle. In our (perturbative) analysis, we only consider the first-order effects of the scalar field, i.e. first-order effects in $Q_{\rm S}$. Hence, we can write
\beq
a(U)^\lambda=\frac{1}{\mu_0}F_{\rm scal}(U)^\lambda\,.
\eeq
[The change in $\mu$ would have a second-order effect on the acceleration.]
Recalling the above result \eqref{psi0_coords}
\begin{eqnarray}
\phi(t,r,\theta,\varphi)&=& r_0^2 f_s(r_0)f_b(r_0)\sum_{\ell m}Y_{\ell m}(\theta,0)\times \nonumber\\
&&\int\frac{d\omega}{2\pi} G_{\ell m\omega}(r,r_0)S_{\ell m\omega} e^{-i\omega t+im\varphi}\,.\qquad
\end{eqnarray}
one can implement (at least numerically) these modifications and compute the acceleration
\be
\label{accscalar}
a(U)^\lambda=
\frac{Q_{\rm S}}{\mu}      P(U)^{\lambda\mu}\nabla_\mu \phi\,.
\ee
The explicit result is quite cumbersome and we refrain from displaying it.

Comparing with the Poynting-Robertson-like description of accelerated motions Eq. \eqref{aU_PR}, whereby
\be
\label{comparisons}
a(U)^\lambda=
\frac{\sigma}{\mu} P(U)^\lambda{}_\mu T^{\mu\nu}_{\rm test\,scal}U_\nu\,,
\ee
apart from $\sigma \leftrightarrow  Q_{\rm S}$, in order to match \eqref{accscalar} and \eqref{comparisons}, one needs
\be
T_{\rm test\,scal}^{\mu\nu}=-P(U)^{\mu \sigma} \nabla_\sigma \phi U^\nu-P(U)^{\nu \sigma} \nabla_\sigma \phi U^\mu\,.
\ee

These results generalize the Poynting-Robertson approach discussed in Section III in an interesting way. In particular the present emphasis  is on the non-singular (smooth horizonless) nature of 5D TSs. This requires different b.c.'s (regularity at the cap) from the 'standard' ingoing conditions at the horizon of BHs.

On the basis of this one could extend our present analysis to include effects due to radiative loss of energy and angular momentum. We plan to address these and related issues in the future.

\section{Discussion}

Let us discuss the main results of our present investigation. First of all we have extended MST approach in BH perturbation theory to Top(ological) Stars, that are smooth, horizonless solutions of 5D Einstein-Maxwell theory. Differently from BHs, the geometry of Top Stars smoothly ends with a `cap' at $r=r_b$ and regularity conditions on the perturbations are to be imposed.    

For simplicity we have considered only neutral scalar perturbations, paving the way for
further investigation to include also metric and vector perturbations \cite{Bena:2024hoh, Dima:2024cok}. 
We have tackled the problem within different approaches: MST, PN and JWKB and qSW/AGT
in order to address different aspects of the scalar field dynamics. This is a major accomplishment of the present work. Noticeably this is a first self-force study in a 5D context which incorporates and capitalizes on the 4D experience.
Moreover in order to compare with (Schwarzschild) BHs,  we have mostly considered $P_y=0$ in the KK direction. It would be interesting to extend our present study to the case of charged perturbations, that correspond to wound strings around the KK direction \cite{Cipriani:2024ygw}, possibly with $P_y\neq 0$. 

Before studying the wave equation, we have considered test field perturbations and associated test particle motion. This has later on allowed us to fully determine scalar self-force effects on (quasi) circular orbits.  
We have also briefly outlined how to include radiation losses in this 5D context.

\section*{Acknowledgements}
We would like to especially thank F.~Fucito and F.~Morales for useful discussions and for sharing with us their Mathematica codes. We also acknowledge fruitful scientific exchange with A.~Cipriani, T.~Damour, G.~Dibitetto,  C.~Gambino, A.~Geralico, A.~Grillo, C.~Kavanagh, A.~Ottewill, P.~Pani, H.~Poghosyan, F.~Riccioni, M.~Sasaki, G.~Sudano, D.~Usseglio, B.~Wardell. 
D.~B. acknowledges sponsorship of the Italian Gruppo Nazionale per la Fisica Matematica
(GNFM) of the Istituto Nazionale di Alta Matematica (INDAM).
M.~B. and G.~D.~R. thank the MIUR PRIN contract 2020KR4KN2 \lq\lq String Theory as a bridge
between Gauge Theories and Quantum Gravity'' and the INFN project ST\&FI \lq\lq String Theory and
Fundamental Interactions'' for partial support.

\appendix

\section{Quantum Seiberg-Witten (SW) and Alday-Gaiotto-Tachikawa (AGT) approach}

An alternative approach to the resolution of CHEs governing BH and fuzzball perturbations is based on ${\cal N} = 2$ Supersymmetric Yang-Mills (SYM) theories or equivalently on Liouville conformal field theory (CFT) \cite{Bianchi:2021xpr,Bianchi:2021mft,Consoli:2022eey,Bianchi:2023rlt,Bianchi:2022qph,DiRusso:2024hmd,Aminov:2020yma,
Aminov:2023jve,Bautista:2023sdf,Bonelli:2022ten,Bonelli:2021uvf}.

As originally shown by Seiberg and Witten (SW) \cite{Seiberg:1994aj,Seiberg:1994rs,Nekrasov:2009rc,Nekrasov:2002qd}
 the non-perturbative dynamics of ${\cal N} = 2$ SYM theories in the Coulomb branch, where only the complex scalars in vector multiplets get non-zero vacuum expectation values (VEV), is coded in complex curves whose periods $a_i$ and $\tilde{a}^i$ determine the analytic prepotential 
\be{\cal F}(a_i, m_f, q_a) = {\cal F}_{\rm tree} + {\cal F}_{\rm 1-loop} + {\cal F}_{\rm inst}\,,
\ee
where $q_a$ are the instanton counting parameters, determined by the Renormalization-Group invariant scales $\Lambda_a$ according to $q_a=\Lambda_a^{\beta_a}$ with $\beta_a= 2N_a - N_{f_a}$ the one-loop $\beta$-function coefficients and $m_{f_a}$ the masses of the hypermultiplets in the fundamental representation of the gauge group\footnote{In general, hypermultiplets can belong to any (pseudo)real represantion.} $G=\prod_a G_a$. 

In the simplest case of a single $SU(2)$ gauge group, the SW curve turns out to be a genus-one Riemann surface (torus) a.k.a. an `elliptic curve'. If SYM is coupled to $N_f=(N_L, N_R)$ hypermultiplets in the fundamental (doublet) representation with masses $m_f$, the SW curve reads
\bea
&& q y (x-m_1) (x- m_2) + (x-e_1)(x-e_2) \nonumber\\
&&\qquad  +y^{-1} (x-m_3) (x- m_4) = 0\,, 
\eea
where $q=\Lambda^\beta$ is the instanton-counting parameter with $\beta=2N_c-N_f=4-N_f$ and $N_c=2$ is the number of `colours' and $e_1=-e_2$ related to the scalar VEV $a$. A geometric way of describing the system is offered by the Hanany-Witten brane setup \cite{Hanany:1996ie}. Following this approach one can `decouple' a massive hyper by taking the double scaling limit $m_{N_f} \rightarrow \infty$, $q_{N_f}\rightarrow 0$ with 
$q_{N_f{-}1} =  m_{N_f} q_{N_f}$ fixed.

In order to regulate large volume (IR) divergences, it is convenient to promote the complex variables $x$ and $y$ to non-commuting variables (differential operators) by considering the ${\cal N} = 2$ SYM in a non-commutative $\Omega$ background, 
whereby the 4 dimensional (Euclidean) coordinates $X^\mu$ satisfy $[X_1,X_2]=\varepsilon_1$ and $[X_3,X_4]=\varepsilon_2$,
with $\varepsilon_1, \varepsilon_2$ arbitrary constants. For our purposes it is enough to consider a Nekrasov-Shatasvili (NS) background with $\varepsilon_1 = \hbar, \varepsilon_2 = 0$ in such a way that\footnote{The constant $\hbar$ is only an artificial device to generate non-commutativity and keep the correct dimensions. It is not related to Planck's quantum of action since the resulting wave equation is classical.} 
\be  
\hat{x}=\hbar y\partial_y\,, \qquad \hat{y}=y\,.
\ee
The resulting `quantum' SW (qSW) curve\footnote{Eventually, we will set $\hbar =1$ for notational simplicity, and write the relevant \lq\lq dictionaries" for `classical' perturbations.} turns out to be a second order ordinary linear homogeneous differential equation of Fuchsian type with up to 4 regular singularities. 
\be
{\cal L}_{\rm qSW} U(y)=0\,.
\ee
The case $N_f=4$ ($\beta_{\rm SYM}=0$) has been shown to capture perturbations around Kerr-Newman BHs in asymptotically (A)dS$_4$ which are governed by a Heun Equation with 4 regular singularities. 

Decoupling one massive hyper, as described above, {\it i.e.} passing from $N_f=4$ to $N_f=3$, one finds a Confluent Heun Equation (CHE) with two regular and one irregular singularities, 
\be\label{CHE}
H''(\xi)+\left({\gamma\over \xi}+{\delta\over \xi-1}+\eta\right)H'(\xi)+{\alpha\xi-\beta\over \xi(\xi-1)}H(\xi)=0\,,
\ee
where we used $\xi$ to avoid confusion with the variables used above. Eq. \eqref{CHE}   can be put in its normal form by defining 
\be
H(\xi)=e^{-\eta \xi/2}\xi^{-\gamma/2}(1-\xi)^{-\delta/2}\Psi(\xi)\,,
\ee
so that
\be
\Psi''(\xi)+Q_{\rm CHE}(\xi)\Psi(\xi)=0\,,
\ee
and
\bea\label{QCHE}
Q_{\rm CHE}(\xi)&{=}&{\gamma(2{-}\gamma)\over 4\xi^2}{+}{\delta(2{-}\delta)\over 4(1{-}\xi)^2}{-}{\eta^2\over 4}\nonumber\\
&{+}&{2\beta{+}\gamma\delta{-}\gamma\eta\over 2\xi}{+} {2\beta{-}2\alpha{+}\gamma\delta{+}\delta\eta\over 2(1{-}\xi) }\,.
\eea
As already observed, perturbations around the most general (non-extremal) KN BH is D=4 can be brought to this form with the CHE parameters related to the mass $M$, charge $Q$, angular momentum $J=M a_J$ of the BH and to the energy $\omega$, angular momentum $\ell$, azymuthal number $m$, charge $q$ and mass $\mu$ of the perturbation. In fact, as shown in \cite{Bianchi:2021xpr,Bianchi:2021mft,Consoli:2022eey,Bianchi:2023rlt,Bianchi:2022qph,DiRusso:2024hmd} 
all integrable linear perturbations of higher dimensional branes, fuzzballs and BHs can be related to Heun Equation, up to further confluences and reductions.  
 
Comparing CHE with the qSW curve (with $\hbar=1$) for $SU(2)$ SYM with $N_f=3=(1,2)$ hypers  
\bea
&&Q_{1,2}(\xi){=}{-}{q^2\over 4}{+}{1{-}(m_1{+}m_2)^2\over 4(1{-}\xi)^2}{+} {1{-}(m_1{-}m_2)^2\over 4\xi^2}\nonumber\\
&{+}& \frac{1{-}2(m_1^2{+}m_2^2){-}2q(1{-}m_1{-}m_2){+}4u}{4\xi(1{-}\xi)}{+} \frac{qm_3}{\xi}\,,
\eea
with $1,2$ referring to the brane configuration in the Hanany-Witten setup.
Although this is not symmetric in $m_1$, $m_2$ and $m_3$ all the `observables' are symmetric. 
Moreover $u = \langle {\rm Tr} \phi^2 \rangle = a^2 + ...$ is the gauge-invariant Coulomb branch parameter and one can deduce the following dictionary 
\bea
m_1&=&{\gamma+\delta-2\over2},\nonumber\\
m_2&=&{\delta-\gamma\over2},\nonumber\\
m_3&=& {\alpha\over \eta} - {\gamma+\delta\over 2},\nonumber\\
u&=&-\alpha+\beta+\left({\gamma+\delta-1\over 2}\right)^2+\eta,\nonumber\\
q&=&\eta\,.
\eea

In order to determine the quantum SW period $a$ it proves convenient to consider the opposite quantization rule 
\be  
\hat{x}= x \quad , \quad \hat{y}=e^{- \hbar \partial_x}
\ee
that leads to the `finite difference' equation 
\be
{\cal Y}(x) {\cal Y}(x-1) - P(x)  {\cal Y}(x-1) + q M(x) =0\,, 
\ee
with 
\be
{\cal Y}(x) = P_L\left(x+{1\over 2}\right) y(x) = \left(x-m_3+{1\over 2}\right)  y(x)
\ee
and 
\bea
&&P(x) = x^2-u+q \left(x+\frac12 -m_1-m_2-m_3\right)\,,\nonumber\\
&& qM(x) = Q(x)\nonumber\\ 
&&=q \left(x-\frac12 -m_1\right)\left(x-\frac12 -m_2\right)\left(x-\frac12 -m_3\right)\,.\nonumber\\
\eea
Integrability imposes a condition on $u=a^2+...$ or the quantum SW period $a = a(u,q,m_f, \hbar)$  
that can be written in terms of continuous fractions \cite{Poghosyan:2020zzg}

\bea
\label{rec_rel_PQ}
P(a)&=&\frac{Q(a)}{P(a-1)-\frac{Q(a-1)}{P(a-2)-\ldots} }\nonumber\\
&-&
\frac{Q(a+1)}{P(a+1)-\frac{Q(a+2)}{P(a+2)-\ldots }}\,,
\eea
and solved recursively~\footnote{There is a sign ambiguity in the choice of $\sqrt{u}$ and thus in $a$.}
\beq
a=\sqrt{u}+ {q} a_1(u)+{q}^2 a_2(u)+O({q}^3)\,.
\eeq
Setting $\Sigma_1 = m_1+m_2+m_3$, $\Sigma_2=m_1m_2+m_1m_3+m_2m_3$, $\Pi =m_1m_2m_3$, the first three terms of the expansion read\footnote{The same results can be obtained imposing the quantum Matone relation~\cite{Matone:1995rx,Flume:2004rp} $u=-q\partial_q{\cal F}$ and adding the one-loop term.}

\begin{widetext}
\be\label{a6inst}
a_1=\frac{1}{4\sqrt{u}}\left(\Sigma_1-1-{4\Pi\over 1-4u}\right)
\ee
\bea
a_2&=&  -\frac{ (3 u-2)}{64 u^{3/2}(-1+u)}\Sigma_1^2
-\frac{(36 u^2-29 u+2)}{8(-1+4 u)^2 u^{3/2}(-1+u)} \Sigma_1 \Pi+\frac{1}{16 u^{3/2}} \Sigma_1 
+\frac{3}{16 u^{1/2}(-1+4 u) (-1+u)} \Sigma_2^2\nonumber\\
&+&\frac{1}{ 32 u^{1/2}(-1+u)} \Sigma_2
-\frac{ (60 u^2-35 u+2)}{4 (-1+u) u^{3/2} (-1+4 u)^3} \Pi^2
+\frac{ (12 u-1)}{4 (-1+4 u)^2 u^{3/2}} \Pi
-\frac{ (4 u^2+3 u-8)}{ 256 u^{3/2}(-1+u)}\nonumber\\
\eea
\bea
a_3&=& \frac{ (-18+413 u-4705 u^2+15260 u^3-18480 u^4+6720 u^5)}{4 u^{5/2} (-1+u)^2 (-9+4 u) (-1+4 u)^5} \Pi^3\nonumber\\
&+&\left[\frac{(-54+1131 u-11695 u^2+33348 u^3-35280 u^4+11200 u^5)}{16 (-1+4 u)^4 (-9+4 u) (-1+u)^2 u^{5/2}} \Sigma_1
-\frac{3 (560 u^4-840 u^3+371 u^2-39 u+2)}{16 (-1+4 u)^4 (-1+u)^2 u^{5/2}}\right]\Pi^2\nonumber\\
&+&\left[\frac{ (4800 u^5-16720 u^4+17876 u^3-7447 u^2+915 u-54)}{64 (-1+4 u)^3 (-9+4 u) (-1+u)^2 u^{5/2}} \Sigma_1^2
-\frac{(720 u^4-1280 u^3+703 u^2-95 u+6)}{32 (-1+4 u)^3 (-1+u)^2 u^{5/2}} \Sigma_1\right.\nonumber\\
&-& \frac{ (-27+497 u-1120 u^2+560 u^3)}{16 u^{3/2}(-1+4 u)^3 (-9+4 u) (-1+u)^2} \Sigma_2^2
-\frac{ (240 u^3-440 u^2+179 u-9)}{32 u^{3/2}(-1+4 u)^2 (-9+4 u) (-1+u)^2} \Sigma_2 \nonumber\\
&-&\left. \frac{(768 u^6-7040 u^5+29760 u^4-44376 u^3+23945 u^2-3363 u+216)}{256 (-1+4 u)^3 (-9+4 u) (-1+u)^2 u^{5/2}} \right]\Pi\nonumber\\
&+&\frac{(5 u^2-5 u+2)}{256 (-1+u)^2 u^{5/2}} \Sigma_1^3
-\frac{(9 u^2-13 u+6)}{256 (-1+u)^2 u^{5/2}} \Sigma_1^2\nonumber\\
&+&\left[-\frac{3 (1-15 u+20 u^2)}{64 u^{3/2}(-1+4 u)^2 (-1+u)^2}\Sigma_2^2
-\frac{(3 u-1)}{ 128 u^{3/2}(-1+u)^2}\Sigma_2
+\frac{ (4 u^3+13 u^2-43 u+24)}{1024 (-1+u)^2 u^{5/2}}\right]\Sigma_1\nonumber\\
&+&\frac{3 (1-15 u+20 u^2)}{ 64 u^{3/2} (-1+4 u)^2 (-1+u)^2}\Sigma_2^2
+\frac{ (3 u-1)}{ 128 u^{3/2}(-1+u)^2}\Sigma_2
-\frac{ (4 u^3-3 u^2-11 u+8)}{ 1024 (-1+u)^2 u^{5/2}}\,.
\eea
Denoting $\Pi=\Sigma_3$ to make the notation uniform,  we can write formally
\bea\label{a6inst_formal}
a_1&=&c_{(1)0}(u) +c_{(1)i}(u)\Sigma_i\,,\nn\\
a_2&=&c_{(2)0}(u) +c_{(2)i}(u)\Sigma_i +c_{(2)ij}(u)\Sigma_i\Sigma_j\,,\nonumber\\
a_3&=&c_{(3)0}(u) +c_{(3)i}(u)\Sigma_i+c_{(3)ij}(u)\Sigma_i\Sigma_j +c_{(3)ijk}(u)\Sigma_i\Sigma_j\Sigma_k\,,
\eea
identifying the $\Sigma_i$ structure of these relations.

\end{widetext}
\begin{table}
\caption{\label{tab:table1coefs} Explicit expressions for the first coefficients entering Eq. \eqref{a6inst_formal}.}
\begin{ruledtabular}
\begin{tabular}{ll}
$c_{(1)0}(u)$&$  -\frac{1}{4\sqrt{u}}$\\
$c_{(1)1}(u)$&$ \frac{1}{4\sqrt{u}}$\\
$c_{(1)2}(u)$&$0$\\
$c_{(1)3}(u)$&$ -\frac{1}{\sqrt{u}(1-4u)}$\\
\hline
$c_{(2)11}(u)$&$    -\frac{ (3 u-2)}{64 u^{3/2}(-1+u)}$\\
$c_{(2)13}(u)$&$ -\frac12 \frac{(36 u^2-29 u+2)}{8(-1+4 u)^2 u^{3/2}(-1+u)}$\\
$c_{(2)22}(u)$&$ \frac{3}{16 u^{1/2}(-1+4 u) (-1+u)}$\\
$c_{(2)33}(u)$&$ -\frac{ (60 u^2-35 u+2)}{4 (-1+u) u^{3/2} (-1+4 u)^3}$\\
$c_{(2)1}(u)$&$ \frac{1}{16 u^{3/2}}$\\
$c_{(2)2}(u)$&$  \frac{1}{ 32 u^{1/2}(-1+u)}$\\ 
$c_{(2)3}(u)$&$  \frac{ (12 u-1)}{4 (-1+4 u)^2 u^{3/2}}$\\ 
$c_{(2)0}(u)$&$   -\frac{ (4 u^2+3 u-8)}{ 256 u^{3/2}(-1+u)}$\\
\end{tabular}
\end{ruledtabular}
\end{table}

Quite remarkably $a=\ell +{1\over 2} + ... $ is exactly related to the `renormalized' angular momentum $\nu = \ell + ...$ introduced by MST in order to express solutions of the CHE in terms of (confluent) hypergeometric functions by 
\be a=\nu+{1\over 2}\, .
\ee
We have checked this relation up to order 12 in the expansion in powers of $q$. 

The last step in the qSW approach consists in the `combinatorial'  reconstruction of the wavefunction, relying on the AGT correspondence (after Alday, Gaiotto and Tachikawa, \cite{Alday:2009aq}) between ${\cal N}=2$ SYM in D=4 with $G=SU(2)$ and Liouville CFT in D=2. Without entering too much in the details of the AGT correspondence, let us mention that conformal blocks of primary operators are related to ratios of instanton  partition functions. Conformal blocks satisfy conformal Ward identities a.k.a. BPZ equations (after Belavin, Polyakov and Zamolodchikov, \cite{Belavin:1984vu}) 
corresponding to insertions of null fields. In the special cases of interest for us, these equations are exactly of the Heun type and its associated confluences.  

Parametrizing the central charge $c$ of the Virasoro algebra and the background charge $Q$ of Liouville CFT \cite{Belavin:1984vu} as 
\be
c = 1 + 6Q^2\,, \qquad Q = b + {1\over b}\,,
\ee
the relation with the gauge theory in the non-commutative $\Omega$ background is 
\be
b= \sqrt{\varepsilon_2\over \varepsilon_1}\,,
\ee
so that the NS limit corresponds to the `semiclassical' limit\footnote{In this limit the central charge $c$ and background charge $Q$ are very large.} $b\rightarrow 0$. Moreover primary fields with `chiral' vertex operators  
\be
V_P= e^{2({Q\over 2} - P)\varphi}\,, 
\ee
with $P= {p\over b}$ have scaling dimension $\Delta(p) = ({Q\over 2} - P) ({Q\over 2} + P)= {Q^2\over 4} - {p^2\over b^2}$. The primary field of type [2,1] with `momentum' $p_d= b^2 + {1\over 2}$ and $\Delta_d = -{3\over 4} b^2 - {1\over 2}$ is `degenerate' in that it has a null descendant at level 2. For later use, let us define the shifted `momenta'
\be
p(\alpha) = p +  {1\over 2} \alpha b^2\,.
\ee
A five-point conformal block $\Psi(z)$ with insertion of a degenerate field in the point $z$, four primaries at $0$, $z_2$, $1$ and $\infty$ satisfies a HE in the variable $z$. We are interested in the confluence $z_2\rightarrow 0$ with $k_2, p_3 \rightarrow \infty$ while $w= (k_2-p_3)z_2$ and ${k}=k_2+k_3$ are kept finite. The resulting BPZ equation is of the CHE form and can be put in canonical form with $z=z_{\rm cft}$  
 \bea\label{QCFT}
Q^{(2,1)}_{\rm cft}(z)&=&- {w^2 \over 4z^4}+{w {k} \over z^3}+{U-{1\over 4} + w ({1\over 2} - {k}_0) \over z^2(z-1)}\nonumber\\
&+&{{1\over 4} - {k}_0^2 \over z(z-1)^2} +{{1\over 4} - {p}_0^2 \over z(z-1)}\,,
\eea
where, as before,  $1,2$ refer to the brane configuration in the Hanany-Witten setup.
Here $z=q_1$, $w=q_1 q_2$, ${k}=-m_3$, $U= u$, ${k}_0= {m_1+m_2 \over 2}$ and ${p}_0= {m_1-m_2 \over 2}$ are related to the Liouville `momenta' of the primary fields. Moreover $\varepsilon_1=1=\hbar$, $b^2 = \varepsilon_2/\varepsilon_1 = \varepsilon_2$, $p^{\pm} = p \pm {b^2\over 2}$ and ${k}_{d}= - m_{\rm bi-fund} = b^2+{1\over 2}=p_d$ is the momentum of the `degenerate' conformal field having a null descendant at level 2.
In the `semiclassical' limit $b\rightarrow 0$, the two independent conformal blocks are
\begin{widetext}
\beq\label{z41nonexp}
{\Psi}_\pm(z) = \lim_{b\rightarrow 0}  e^{w\over 2z} z^{\pm a +{1\over 2}} 
(1-z)^{(2{k}_0 - 1 - b^2)(2{k}_{d} + 1 + b^2)\over 2b^2}  
{Z^{\rm inst}_{{p}_0, {k}_0, a^\mp, {k}_{d}, a, {k}} (z, {w\over z}) \over  
Z^{\rm inst}_{{p}_0, {k}_0, a, {k}} (z)}\,,
\eeq 
where $a^\pm = a \pm {b^2\over 2}$, $Z^{\rm inst}_{{p}_0, {k}_0, a, {k}} (z)$ is the instanton partition function for a single $SU(2)$ SYM theory with $q_1=z$, while $Z^{\rm inst}_{{p}_0, {k}_0, a^\mp, {k}_{d}, a, {k}} (z, {w\over z})$ is the instanton partition function for a quiver SYM theory with $SU(2)\times SU(2)$ group with $q_1=z$ and $q_2={w\over z}$.
The former can be expressed as a sum over pairs of Yang tableaux $\vec{\cal Y}$ with $|\vec{\cal Y}|$ boxes in total 
\be
Z^{\rm inst}_{{p}_0, {k}_0, a, {k}} (z) = \sum_{\vec{\cal Y}} z^{|\vec{\cal Y}|} 
{{\cal Z}^{\rm bi-fund}_{\emptyset,\vec{\cal Y}}({p}_0, a, -{k}_0) {\cal Z}^{\rm fund}_{\vec{\cal Y}}(a, -{k}) \over  {\cal Z}^{\rm bi-fund}_{\vec{\cal Y},\vec{\cal Y}}(a, a, {1+b^2 \over 2}) }
\ee
The latter  can be expressed as a sum over two pairs Yang tableaux $\vec{\cal Y}_{1,2}$ with $|\vec{\cal Y}_{1,2}|$ boxes
 \be
Z^{\rm inst}_{{p}_0, {k}_0, a^\mp, {k}_{d}, a, {k}} \left(z, {w\over z} \right) = \sum_{\vec{\cal Y}_{1,2}} z^{|\vec{\cal Y}_1|} \left({w\over z}\right)^{|\vec{\cal Y}_2|}
{{\cal Z}^{\rm bi-fund}_{\emptyset,\vec{\cal Y}_1}({p}_0, a^\mp, -{k}_0) 
{\cal Z}^{\rm bi-fund}_{\vec{\cal Y}_1,\vec{\cal Y}_2}({{a^\mp}}, a, -{k}_{d})
{\cal Z}^{\rm fund}_{\vec{\cal Y}_2}(a, -{k}) \over  {\cal Z}^{\rm bi-fund}_{\vec{\cal Y}_1,\vec{\cal Y}_1 }(\mp a, \mp a, {1+b^2 \over 2}) {\cal Z}^{\rm bi-fund}_{\vec{\cal Y}_2,\vec{\cal Y}_2 }(a, a, {1+b^2 \over 2}) }
\ee
In turn 
\bea
{\cal Z}^{\rm bi-fund}_{\vec{\cal Y},\vec{\cal Y}'}(p, p',m)&=& \prod^{1,2}_{\sigma, \sigma'} \prod_{s\in {\cal Y}_\sigma} \left[{\cal E}_{{\cal Y}_\sigma, {\cal Y}'_{\sigma'}} (\sigma p - \sigma' p', s) - m  \right]
\prod_{t\in {\cal Y}'_{\sigma'}}\left[-{\cal E}_{{\cal Y}_\sigma, {\cal Y}'_{\sigma'}} (\sigma' p' - \sigma p, t) - m  \right]\nonumber\\
{\cal Z}^{\rm fund}_{\vec{\cal Y}}(a,m)&=& \prod^{1,2}_{\sigma} \prod_{s\in {\cal Y}_\sigma} \left[-{\cal E}_{{\cal Y}_\sigma, \emptyset} (\sigma a, s) + m  \right]\,,
\eea
where 
\be
{\cal E}_{{\cal Y}, {\cal Y}'}(p,s_{h,j}) =  p - [\rho^t_{{\cal Y}'}(j) - h] + b^2 [\rho_{{\cal Y}}(h) - j +1] - {1+b^2\over 2}\,,
\ee
with $\rho_{{\cal Y}}(h)$ the number of boxes in the $h^{\rm th}$ row of ${\cal Y}$ and $\rho^t_{{\cal Y}'}(j)$  the number of boxes in the $j^{\rm th}$ column of ${\cal Y}'$.
Though somewhat cumbersome to describe, the procedure can be easily  implemented  by using an algebraic manipulator.

For instance, the simplest case of a pair `empty' Young tableaux $(\bullet, \bullet)$ for which 
$\rho_{\bullet}(h)=0=\rho_{\bullet}^t(j)$ one has ${\cal E}_{(\bullet, \bullet)}(p,s=(1,1)) = p + {1\over 2} - {b^2\over 2} $.

The next possibility are the pairs $(\bullet, \Box)$ and $(\Box, \bullet)$ for which 
$\rho_{\Box}(h)=1=\rho_{\Box}^t(j)$ one has
${\cal E}_{(\bullet, \Box)}(p,s=(1,1)) = p - {1\over 2} - {b^2\over 2} $ and
${\cal E}_{(\Box, \bullet)}(p,s=(1,1)) = p + {1\over 2} + {b^2\over 2} $

In the limit $b\rightarrow 0$ both numerator and denominator in \eqref{z41nonexp} diverge but their ratio $\Psi$ remains finite. Using the `combinatorial' Yang-tableaux construction one finds a double expansion in $z=q_1$ and ${w\over z} = q_2$ (so that $w=q_1q_2$), whose first few terms read 
\be\label{z54pm}
\Psi_\pm (z) =  z^{\pm a + {1\over 2}}\left\{ 1 + z {a^2- {1\over 4} + {k}_0^2 - {p}_0^2 \over 1 \pm 2 a}  - {w \over z} {{k} \over (1 \mp 2 a)}  - w {{k}[   
4(3\pm 2a)({k}_0^2-{p}_0^2) - (3\mp 2a) (1\pm 2a)^2]\over 4(1\mp 2a) (1\pm 2a)^2 } + ... \right\}
\ee
with monodromy (`Floquet' exponent)  $\Psi_\pm (e^{2\pi i} z) = - e^{\pm 2\pi i a} \Psi_\pm (z)$.
More terms of the simultaneous expansion in $q_1$ and $q_2$ are displayed in Appendix \ref{appc}.

Quite remarkably the procedure does not require the explicit expression of the qSW period $a$ but its consistency does, very much as for $\nu$ in the MST approach. However, contrary to the latter approach, the dependence on $\ell \sim U\sim u$ remains implicit in $a$ while $\ell \sim \nu$, as we have seen, appears explicitly in the MST recursion relations. 

In the two limits $q_2= {w\over z} \ll z = q_1$ and $q_2= {w\over z} \gg z = q_1$ the `instanton' sums boil down to (confluent) hypergeometric functions
\[
\begin{array}{lll}
 z\gg {w\over z}:  &   \Psi^{\rm in}_{\pm}(z) &= \lim_{{w\over z} \rightarrow 0} \Psi_{\pm}(z) \cr 
 & &  
=z^{\pm a + {1\over 2}} (1-z)^{{k}_0 + {1\over 2}} {}_2F_1(\pm a + {k}_0+{p}_0+ {1\over 2}, \pm a + {k}_0-{p}_0+ {1\over 2}; \pm 2 a + 1; z)\,, \cr
& \cr
 z\ll {w\over z}:  &    \Psi^{\rm up}_{\pm}\left({w\over z}\right)&= \lim_{z \rightarrow 0} \Psi_{\pm}(z) \cr
&&=  
\left({w\over z}\right)^{\pm a + {1\over 2}} e^{-{w\over 2z}}{}_1F_1(\mp a - {k} - {1\over 2}, \mp 2 a + 1; {w\over z})\,, \cr
\end{array}
\]
\end{widetext}
that can be expressed in terms of the variables $1-z$ and $z/w$ using the connection formulae for (confluent) hypergeometric functions, {\it i.e.} `fusing' $F_{\alpha\beta}$ and `braiding' $B_{\alpha\beta}$ matrices (with $\alpha, \beta= \pm  1)$ for Liouville conformal blocks, {\it viz.}
\bea 
\Psi^{\rm in}_{\alpha}(z) &=& \sum_\beta F_{\alpha\beta} \widetilde\Psi^{\rm in}_{\beta}(1-z)\,,\nonumber\\
\Psi^{\rm up}_{\alpha}({w\over z}) &=& \sum_\beta B_{\alpha\beta} \widetilde\Psi^{\rm up}_{\beta}({z\over w})\,, 
\eea
with
\bea
\widetilde\Psi^{\rm in}_{\pm}(1-z) &=&(z)^{a +{1\over 2}} (1-z)^{\pm {k}_0+{1\over 2}}{}_2F_1^\pm \nonumber\\  
\widetilde\Psi^{\rm up}_{\pm}({z\over w})  &=& ({z\over w})^{\pm {k} +{1\over 2}} e^{\pm {w\over 2 z}}{}_2F_0^\pm \,,
\eea
with
\bea
{}_2F_1^\pm &=& {}_2F_{1}(a\pm {k}_0+{p}_0 +{1\over 2} , a\pm {k}_0-{p}_0 +{1\over 2};\nonumber\\
&&  1\pm 2{k}_0; 1-z)\,,\nonumber\\
{}_2F_0^\pm &=& {}_2F_{0}(a\pm {k} +{1\over 2} , - a \pm {k} +{1\over 2}; \pm {z\over w})\,,
\eea
and
\bea
F_{\alpha,\beta} &=& {\Gamma(1+2\alpha a) \Gamma(-2\beta {k}_0)  \over  
\Gamma({1\over 2} + {p}_0 + \alpha a - \beta {k}_0) \Gamma({1\over 2} - {p}_0 + \alpha a - \beta {k}_0)}\,,\nonumber\\
F^{-1}_{\alpha,\beta} &=& {\Gamma(1+2\alpha {k}_0) \Gamma(-2\beta a)   \over  
\Gamma({1\over 2} + {p}_0 + \alpha {k}_0  - \beta a) \Gamma({1\over 2} - {p}_0 + \alpha {k}_0  - \beta a)}\,,\nonumber\\
B_{\alpha,\beta} &=& \beta^{-{1\over 2} + \alpha a + \beta {k}}
{\Gamma(1-2\alpha a) \over  
\Gamma({1\over 2} - \alpha a - \beta {k})}\,,
\eea
with $\alpha, \beta = \pm$.

\subsection{CHE for TS and `comparison' MST vs qSW/AGT}

In order to compare the MST approach with the qSW/AGT approach 
\cite{Bianchi:2021xpr,Bianchi:2021mft,Bonelli:2021uvf,Bonelli:2022ten,Consoli:2022eey}, 
it is convenient to introduce a third variable\footnote{In \cite{Fucito:2023afe}, $z= {r_+ - r_- \over r - r_-}$ with $r_+$ the outer horizon and $r_-$ the inner horizon (or the singularity $r_-=0$ for Schwarzschild).}
\be
z_{\rm cft} = {r_b - r_s \over r - r_s} = {1\over 1-z_{\rm in}} = -{\epsilon \over z_{\rm up}}
\ee
such that $r=r_b$ (cap) corresponds to $z_{\rm cft}=1$ and $r\rightarrow \infty$ to $z_{\rm cft} = 0$, with 
\be
z_{\rm in} = -{r-r_b\over r_b-r_s} \qquad z_{\rm up} = \omega (r-r_s) 
\ee
such that $r=r_b$ (cap) corresponds to $z_{\rm in}=0$ or $z_{\rm up}=-\epsilon$ and $r\rightarrow \infty$ to $z_{\rm in}, z_{\rm up} \rightarrow \infty$.
Putting the radial equation \eqref{tsw1} with $P_y=0$ in normal form, with
\beq
R(r)={\Psi(r)\over \sqrt{(r-r_s)(r-r_b)}}\,,
\eeq 
the equation satisfied by $\Psi$ is
\beq
\label{CHETS}
\Psi''(r)+Q_{\rm TS}(r)\Psi(r)=0\,, \nn\\
\eeq
where
\bea
\label{QTSr}
Q_{\rm TS}(r)&{=}&\frac{1}{4(r{-}r_s)^2(r{-}r_b)^2}\left(4\omega^2 r^4{-}4r_b\omega^2 r^3{-}4L r^2\right.\nonumber\\
&{+}&\left. 4L(r_b{+}r_s)r{+}r_s^2{+}r_b^2{-}2r_sr_b(1{+}2L)\right)\,,
\eea
with $L=\ell(\ell+1)$, one can map into \eqref{CHETS} in normal form in the $z_{\rm in}$ variable:
\bea
\label{QTSin}
&&Q^{(1,2)}_{\rm TS}({z_{\rm in}}) = (r_b-r_s)^2 Q_{\rm TS}(r) = {1\over 4{z_{\rm in}}^2}\nonumber\\
&+&(r_b-r_s)^2\omega^2+{r_b-r_s-4r_s^3\omega^2\over4(r_b-r_s)(1-{z_{\rm in}})^2}\nonumber\\
&+&{r_b+2Lr_b+4r_s^3\omega^2-6r_s^2r_b\omega^2-r_s-2Lr_s\over 2(r_b-r_s)(1-{z_{\rm in}})} \nn\\
&+& {r_b+2L r_b-r_s-2L r_s-2r_b^3\omega^2\over2(r_b-r_s){z_{\rm in}}}\,.
\eea
where $r = r_b - (r_b-r_s) z_{\rm in}$ and the notation $N_f=(1,2)=(N_L,N_R)$ refers to the brane configuration in the Hanany-Witten setup.
One can also map \eqref{CHETS} in the $z_{\rm up}$ variable:
\bea
\label{QTSup}
Q^{(1,2)}_{\rm TS}({z_{\rm up}}) = (r_b-r_s)^2 Q_{\rm TS}(r)
\eea
where $r = r_s - {r_b-r_s \over \epsilon} z_{\rm up} $ which is in normal form, too.
On the contrary, if one uses $z_{\rm cft}$ one has to re-canonize $\psi$ into $\psi= \tilde\psi(z_{\rm cft})/z_{\rm cft}$ since $r= r_s + {r_b-r_s \over z_{\rm cft}} $ and 
\bea
\label{QTScft}
Q^{(2,1)}_{\rm cft}({z_{\rm cft}})&=& {(r_b-r_s)^2\over z_{\rm cft}^4} Q_{\rm TS}(r)\,.
\eea

Here
\bea
\alpha_{\rm in}&=&\frac{2i\omega(r_b-r_s)}{4(r_s-r_b)} \left( 2r_s(2-2i\omega(r_b-r_s))+{2ir_s^{3/2}\eta\over \sqrt{r_b-r_s}}\right.\nonumber\\
&-&\left. r_b(4+2i\omega(r_b-r_s)) \right)\nonumber\\ 
&=& 2i\epsilon + 2 \epsilon \left( {\tau} - i{\kappa}\right)\nonumber\\
\beta_{\rm in} &=&L+2i\omega(r_b-r_s)\left({1\over2}+{ir_s^{3/2}\over2(r_b-r_s)^{3/2}}\right)\nonumber\\
&-&{r_b^3\omega^2 \over  (r_b-r_s) }\nonumber\\ 
&=& \ell(\ell+1) -i\epsilon +2\epsilon \kappa  - {8\over 27\epsilon} (\epsilon-\tau)^3  \nonumber\\
\gamma_{\rm in} &=& 1\nonumber\\
\delta_{\rm in} 
&=&  1 +\frac{2 r_s^{3/2}\epsilon}{(r_b-r_s)^{3/2}}=1- 2 {\kappa}\nonumber\\ 
\eta_{\rm in} &=& -2i\epsilon = +2i\omega(r_b-r_s)
\eea
with $\epsilon = \omega (r_s-r_b) $, $\kappa = {\omega r_s^{3/2} \over (r_b-r_s)^{1/2}}$ and $\tau = \omega (r_s + {r_b\over 2})$ so that $\omega r_b=-\frac23 (\epsilon-\tau)$ and $\omega r_s=\frac13 (\epsilon+2\tau)$, as before.

One can now use the `combinatorial' Yang-tableaux construction for $\Psi$ and find
\bea
\Psi_\pm &=&  z^{\pm a + {1\over 2}}\left\{ 1 + z {a^2- {1\over 4} + {k}_0^2 - {p}_0^2 \over 1 \pm 2 a}\right. \nonumber\\
&-& {w \over z} {{k} \over (1 \mp 2 a)} \nonumber\\
&-& \left. w {{k}[   
4(3\pm 2a)({k}_0^2-{p}_0^2) - (3\mp 2a) (1\pm 2a)^2]\over 4(1\mp 2a) (1\pm 2a)^2 } + ... \right\}\nonumber\\
\eea
with $z=z_{\rm cft}$ and
\bea\label{dict1}
{k} &=& m_3= - i \tau  \quad , \quad m_1=- m_2 = - \kappa \nonumber\\ 
w_{\rm cft}&=& -q_{TS} = + 2i\omega (r_b-r_s) \nonumber\\
&=& - 2i\epsilon = -\eta_{\rm in}\nonumber\\
u&=&\left(\ell+\frac{1}{2}\right)^2\nonumber\\
&+&\frac{1}{3}\Big[\epsilon(3i-\epsilon+6i \kappa)-4\epsilon\tau-4\tau^2\Big]\,.
\eea
Moreover 
\bea\label{dict2}
U&=&u+ q m_3+q m_{1=2} \nonumber\\
&=& \left(\ell+\frac{1}{2}\right)^2 + i\epsilon (1+2\kappa) - \frac{1}{3}(\epsilon+2\tau)^2 + 2i\epsilon(-i\tau-\kappa)\nonumber\\ 
&=& \left(\ell+\frac{1}{2}\right)^2+ i\epsilon - \frac{1}{3}(\epsilon+2\tau)^2 +2\epsilon \tau \,,
\eea
so that
\be\label{dict3}
{k}_0= {m_1+m_2\over 2} = 0  \quad , \quad {p}_0 = {m_1-m_2\over 2} = -\kappa
\ee 
We find a double expansion, suitable for PN approximation, whose first few terms read 
\bea
\Psi_\pm (z) &{=}&  z^{\pm a{+}{1\over 2}}\left\{ 1{+}z {a^2{-}{1\over 4}{+}\kappa^2 \over 1 \pm 2 a}{+}{2\epsilon \over z} {\tau \over (1 \mp 2 a)} \right. \nonumber\\
&-&\left. 2i\epsilon {-i\tau[   
4(3\pm 2a)(-\kappa^2) - (3\mp 2a) (1\pm 2a)^2]\over 4(1\mp 2a) (1\pm 2a)^2 }\right.\nonumber\\
&&\left. + \cdots \right\}\,,  
\eea
where $z=z_{\rm cft}$, 
to be compared with In and Up solutions or combination thereof in the PN expansion.

Moreover in the two limits $z\equiv {1\over 1-z_{\rm in}} =q_1\ll q_2 = w/z = 2i z_{\rm up}$ and $z=q_1\gg q_2 = w/z$, the expressions simplify to
\bea
\Psi^{\rm up}_{\pm}(\frac{w}{z}=+2iz_{\rm up}) &=& \lim_{z\rightarrow 0} \Psi_{\pm}\left(z,\frac{w}{z}\right)\nonumber\\ 
&=& (+2iz_{\rm up})^{\pm a + {1\over 2}} e^{-iz_{\rm up}}\times \nonumber\\
&& {}_1F_1(\mp a +i\tau - {1\over 2}, \mp 2 a +1; +2iz_{\rm up})\,,\nonumber\\
\eea
which, using 
\be
{}_1F_1(a,b,z) = e^z {}_1F_1(b-a,b,-z)\,,
\ee
are precisely the confluent hypergeometric functions that can be combined to the Tricomi and then Coulomb functions {\it viz.}
\bea 
\Psi^{\rm up}_{\pm}\left(\frac{w}{z}=+2iz_{\rm up}\right) &=& e^{i\varphi} (-2iz_{\rm up})^{\pm a + {1\over 2}} e^{+iz_{\rm up}}\times \nonumber\\ 
&& {}_1F_1(\mp a -i\tau + {1\over 2}, \mp 2 a +1; -2iz_{\rm up})\,,\nonumber\\
\eea
and 
\begin{widetext}
\be
\Psi^{\rm in}_{\pm}\left(z_{\rm cft}={1\over 1- z_{\rm in}}\right) = \lim_{\frac{w}{z}\rightarrow 0} \Psi_{\pm}\left(z,\frac{w}{z}\right) =  z_{\rm cft}^{\pm a +{1\over 2}}(1-z_{\rm cft})^{-\kappa +{1\over 2}} {}_2F_1(\pm a + {1\over 2} -\kappa, \pm a + {1\over 2} -\kappa, \pm 2 a +1; z_{\rm cft})
\ee
using the remarkable relation $a = \nu +{1\over 2}$ 
\bea
\Psi^{\rm in}_{\pm}(z_{\rm cft})= z_{\rm in}^{-\kappa +{1\over 2}} (1-z_{\rm in})^{\pm\nu -\kappa +{1\pm 1\over 2}}
{}_2F_1\left(\pm\nu -\kappa +{1\pm 1\over 2}, \pm\nu +\kappa +{1\pm 1\over 2}, \pm 2 \nu +1 \pm 1; {1\over 1{-}z_{\rm in}}\right)\,.
\eea
Indeed using the identity  for our hypergeometric 
\bea
{}_2F_1(\nu+1-\kappa, -\nu-\kappa; 1; z_{\rm in}) & =& (1-z_{\rm in})^{-\nu-1+\kappa} 
{\Gamma(1) \Gamma(-2\nu-1) \over \Gamma(-\nu-\kappa) \Gamma(-\nu+\kappa)} {}_2F_1(\nu+1-\kappa, \nu+1+\kappa, 2\nu +2; z_{\rm cft})\nonumber\\
\eea
and using $\nu = a- {1\over 2}$ and $2\nu+1=2a$ etc., one finds precisely $\Psi^{\rm in}_{\pm}(z_{\rm cft})$.
Now that we can trust SYM/AGT description we can exploit it for the PN expansion.

In order to match with the PN expansion of the MST-type In solution for $\ell=2$ we need to include a normalization factor.
\be
R_{AGT,in}^{\ell{=}2}(r) {=}\text{Exp}\Big[i\frac{\omega}{2}(r_b{+}r_s){+}\frac{229(r_b{-}r_s)^2\omega^2}{3528}{+}\frac{i(r_b^3{-}3r_br_s^2{+}16r_s^3)\omega^3}{168}\Big](r_b{-}r_s)^{{-}\frac{\omega r_s^{3/2}}{\sqrt{r_b{-}r_s}}} \sqrt{\frac{r{-}r_s}{r{-}r_b}}F^{-1}_{-,-}\Psi_-
\ee
By adding the previous normalization factor the result coming from the two approaches match up to  order $\eta_v^9$ (see \eqref{pnmstl2}). At higher orders in $\eta_v$ one should also take into account the contributions coming from $\Psi_+$ and write the solution in the form 
\be 
R_{AGT,in}(r)= {\cal N} \sqrt{\frac{r{-}r_s}{r{-}r_b}} (F^{-1}_{-,-}\Psi_- + F^{-1}_{-,+}\Psi_+)
\ee
where $N$ is a normalization factor, depending on the physical parameters $\omega$, $r_s$, $r_b$ and $\ell$, necessary since in the AGT approach the overall normalization is fixed while in MST approach it is determined by the sum of the coefficients of the series of hypergeometric functions. Roughly 
\be
\sum_n C_n = {\cal N} 
\ee

\section{MST-type solutions for $\ell=0,1,2$}
In this section we list the first PN terms of the in and up MST-type solutions of the homogeneous radial equation for $\ell=0,1,2$.
In this section we list the first PN terms of the in and up MST-type solutions of the homogeneous radial equation for $\ell=0,1,2$.
We use the notation
\bea
\mathcal{B}&=&\sqrt{4 r_b^4 + 20 r_b^3 r_s + 69 r_b^2 r_s^2 + 74 r_b r_s^3 + 49 r_s^4}\,,\nonumber\\
{\mathcal C}&=& \frac{18r_s^3(r_b+2r_s)}{\mathcal{D}}-1\,,\nonumber\\
{\mathcal D}&=& \mathcal{B}  \left(5 r_b r_s{+}2 r_b^2{+}11 r_s^2\right){+}18 r_s^3 \left(r_b{+}2 r_s\right){+}\mathcal{B} ^2
\,.
\eea
We find
\bea
R_{\rm MST, in}^{\ell=0} &= & -{\mathcal C} + \frac{1}{6} r^2 \eta_v^2 \omega^2 {\mathcal C}   \nn \\ 
& +&\frac{\eta_v^3\omega}{2}\Big[\frac{2r_s^{3/2}}{\sqrt{r_b-r_s}}{\mathcal C}\ln((r_b-r_s)\eta_v^2)+\frac{i \mathcal{B}}{\mathcal{D}}(16r_s^2r_b+7r_b^2r_s+2r_b^3-r_s^3+\mathcal{B}(r_b+r_s))\Big]   \nn\\ 
& +&\frac{r\eta_v^4\omega^2}{120}(100r_s+40r_b-r^3\omega^2){\mathcal C}\nn  \\ 
& -&\frac{r^2\eta_v^5\omega^3}{12}\Big[\frac{2r_s^{3/2}}{\sqrt{r_b-r_s}}{\mathcal C}\ln((r_b-r_s)\eta_v^2)+\frac{i \mathcal{B}}{\mathcal{D}}(16r_s^2r_b+7r_b^2r_s+2r_b^3-r_s^3+\mathcal{B}(r_b+r_s))\Big]\, , 
\eea
\bea
       R_{\rm MST, up}^{\ell=0}&=&-\frac{1}{r\omega}{\mathcal C}-i\eta_v{\mathcal C}+\frac{\eta_v^2}{2r^2\omega}(r^3\omega^2-r_b-r_s){\mathcal C} \nn\\
   &+&\frac{i\eta_v^3}{6r}\Big[{\mathcal C}(r^3\omega^2+3\gamma_E(r_b+2r_s))-36r_s^3\Big] \nn\\
   &-&\eta_v^4{\mathcal D}{\mathcal C}\Big[r^6\omega^4-12r^3(r_b+2r_s)\omega^2(\gamma_E+2\ln(-2ir\omega\eta_v))+8(r_b^2+r_s^2-r_br_s) \nn\\
   &-&\frac{6r^3\omega^2\mathcal{B}(10r_b^3+43r_b^2r_s+100r_br_s^2+123r_s^3+\mathcal{B}(5r_b+9r_s))}{{\mathcal D}{\mathcal C}}\Big] \nn\\
   &-&\frac{i\mathcal{B}\eta_v^5}{120r^2\mathcal{D}}\Big[360r_s^3(r_b+r_s)+20r^3\omega^2(4r_b^3+20r_b^2r_s+47r_br_s^2+37r_s^3+\mathcal{B}(2r_b+5r_s)) \nn\\
   &+&\frac{{\mathcal D}{\mathcal C}}{\mathcal{B}}(r^6\omega^4+30\gamma_E(r_b+2r_s)(r^3\omega^2-r_b-r_s))\Big]\,,
\eea
\bea
R_{\rm MST, in}^{\ell=1}&=&\frac{2 r}{r_b-r_s}-\frac{\eta_v^2 \left(5 \left(r_b+r_s\right)+r^3 \omega ^2\right)}{5 \left(r_b-r_s\right)}+\frac{i r \eta_v^3 \omega  \left(\sqrt{r_b-r_s} \left(r_b+r_s\right)+2 i r_s^{3/2} \ln \left(
   \left(r_b-r_s\right)\eta_v^2\right)\right)}{\left(r_b-r_s\right)^{3/2}} \nn \\
   &{+}&\frac{r^2 \eta_v^4 \omega ^2 \left({-}28 r_b{+}r^3 \omega ^2{-}98 r_s\right)}{140 \left(r_b{-}r_s\right)}{+}\frac{\eta_v^5 \omega  \left(5 \left(r_b{+}r_s\right){+}r^3 \omega ^2\right) \left(2 r_s^{3/2} \ln \left((r_b{-}r_s)\eta_v^2\right){-}i \sqrt{r_b{-}r_s} \left(r_b{+}r_s\right)\right)}{10 \left(r_b{-}r_s\right)^{3/2}} \nn\\
   &+&\frac{r \eta_v^6 \omega ^2}{37800} \Big[\frac{5\omega^4 r^6(r_s-r_b)+15 r^3\omega^2(r_b-r_s)(53r_b+151r_s)-252r_s(17r_b^2+56r_br_s+(50\pi^2-373)r_s^2)}{(r_b-r_s)^2} \nn\\
   &+&\frac{252(32r_b^5+112r_b^4r_s+402r_b^3r_s^2-1343r_b^2r_s^3-1153r_br_s^4+3750r_s^5)}{(4r_b^2+7r_br_s+19r_s^2)^2}-\frac{2520(4r_b^2+7r_br_s+19r_s^2)}{r_b-r_s}\ln(r) \nn\\
   &{+}&2520\left(4r_b{+}11r_s{+}\frac{15r_s^{3/2}(2\sqrt{r_s(r_b{-}r_s)}{-}i(r_b{+}r_s))}{(r_b{-}r_s)^{3/2}}\right)\ln((r_b{-}r_s)\eta_v^2){+}\frac{37800r_s^3}{(r_b{-}r_s)^2}\ln^2((r_b{-}r_s)\eta_v^2)\Big]\, , 
\eea
\bea
R_{\rm MST, up}^{\ell=1}&=&\frac{i}{r^2 \omega ^2}{+}\frac{i \eta_v^2 \left(2 r_b{+}r^3 \omega ^2{+}2 r_s\right)}{2 r^3 \omega ^2}{-}\frac{\eta_v^3 \left(2\omega^2 r^3{+}3(r_b{+}2r_s)(1{-}\gamma_E)\right)}{6 r^2 \omega }\nn\\
&{+}&\frac{i\eta_v^4(36r_b^2{+}48r_br_s{+}36r_s^2{+}20r^3r_b\omega^2{+}40r^3r_s\omega^2{-}5r^6\omega^4)}{40r^4\omega^2} \nn\\
   &+&\frac{\eta_v^5}{60r^3\omega}\Big[2\omega^4r^6-5r^3(r_b+4r_s)\omega^2+15(r_b+2r_s)(\gamma_E \omega^2 r^3+(2\gamma_E-1)(r_b+r_s))\Big] \nn\\
   &{-}&\frac{i\eta_v^6}{3600r^5}\Big[r^3\left((636{-}75\pi^2)r_b^2{+}12(269{-}25\pi^2)r_br_s{+}12(503{-}25\pi^2)r_s^2{+}\frac{22500(r_b{-}r_s)^2r_s^3(r_b{+}2r_s)}{(4r_b^2{+}7r_br_s{+}19r_s^2)^2}\right) \nn\\
   &{-}&\frac{1440(r_b{+}r_s)(2r_b^2{+}r_br_s{+}2r_s^2)}{\omega^2}{-}50r^6(35r_b{+}61r_s)\omega^2{-}25\omega^4r^9{+}60\gamma_E r^3(10r^3(r_b{+}2r_s)\omega^2{-}23r_b^2{-}74r_br_s{-}98r_s^2) \nn\\
   &+&450\gamma_E^2r^3(r_b+2r_s)^2+120r^3(10r^3(r_b+2r_s)\omega^2-4r_b^2-7r_br_s-19r_s^2)\ln(-2ir\omega\eta_v)\Big]\,,
\eea
\bea\label{pnmstl2}
R_{\rm MST, in}^{\ell=2}(r)&{=}&\frac{6r^2}{(r_b{-}r_s)^2}{-}\frac{3 r \eta_v^2 \left(14 \left(r_b{+}r_s\right){+}r^3 \omega
   ^2\right)}{7 \left(r_b{-}r_s\right)^2}{+}\frac{3 i r^2 \eta_v^3 \omega  \left(\sqrt{r_b{-}r_s}
   \left(r_b{+}r_s\right){+}2 i r_s^{3/2} \ln \left(
   \left(r_b{-}r_s\right)\eta_v^2\right)\right)}{\left(r_b{-}r_s\right)^{5/2}} \nn\\
   &+&\frac{\eta_v^4 \left(r^6 \omega ^4-12 r^3 \omega ^2 \left(r_b+8 r_s\right)+84
   \left(4 r_b r_s+r_b^2+r_s^2\right)\right)}{84
   \left(r_b-r_s\right)^2}\nn \\
   &+&\frac{3 r \eta_v^5 \omega  \left(14 \left(r_b+r_s\right)+r^3 \omega
   ^2\right) \left(2 r_s^{3/2} \ln \left(\left(r_b-r_s\right)\eta_v^2\right)-i \sqrt{r_b-r_s}
   \left(r_b+r_s\right)\right)}{14
   \left(r_b-r_s\right)^{5/2}} \nn\\
   &+&\frac{r^2\omega^2 \eta_v^6}{194040(r_b-r_s)^{7/2}}\Big[\sqrt{r_b-r_s}(35r^6\omega^4(r_s-r_b)+2310r^3\omega^2(r_b-r_s)(2r_b+7r_s)\nn\\
   &{-}&132((1470\pi^2{-}11821)r_s^3{+}372r_s^2r_b{+}510r_sr_b^2{-}86r_b^3)){-}5544(r_b{-}r_s)^{3/2}(16r_b^2{+}31r_br_s{+}79r_s^2)\ln(r)\nn\\
   &+&5544(r_b-r_s)((16r_b^2+31r_br_s+79r_s^2)\sqrt{r_b-r_s}-105ir_s^{3/2}(r_b+r_s))\ln((r_b-r_s)\eta_v^2)+\nn\\
   &+&582120\sqrt{r_b-r_s}r_s^3\ln^2((r_b-r_s)\eta_v^2)\Big]\,,
\eea
\bea
R_{\rm MST, up}^{\ell=2}(r)&=&{-}\frac{3}{\omega^3 r^3}{-}\frac{\eta_v^2 \left(9 \left(r_b{+}r_s\right){+}r^3 \omega ^2\right)}{2 r^4 \omega ^3}{+}\frac{3 i (2 \gamma_E {-}3) \eta_v^3 \left(r_b{+}2 r_s\right)}{4 r^3 \omega ^2}\nn\\
&{+}&\frac{\eta_v^4}{56r^5\omega^3}(7r^6\omega^4+14r^3(4r_b+7r_s)\omega^2+144(2r_b^2+3r_br_s+2r_s^2))\nn\\
&+&\frac{i\eta_v^5}{120r^4\omega^2}\Big[-8r^6\omega^4+15(2\gamma_E-3)(r_b+2r_s)(9(r_b+r_s)+r^3\omega^2)\Big]\nn\\
&{-}&\frac{\eta_v^6}{11760 r^6\omega^3}\Big[-245r^9\omega^6+245r^6(7r_b+17r_s)\omega^4{+}12600(r_b+r_s)(5r_b^2+4r_br_s+5r_s^2)\nn\\
&+&r^3\omega^2((735\pi^2-9463)r_b^2+4(735\pi^2-12982)r_br_s{+}4(735\pi^2-18793)r_s^2)\nn\\
&{+}&42\gamma_E r^3\omega^2(379r_b^2+1384r_br_s+1576r_s^2)-4410\gamma_E^2r^3\omega^2(r_b+2r_s)^2\nn\\
&{+}&168r^3\omega^2(16r_b^2+31r_br_s+79r_s^2)\ln(-2ir\eta_v \omega)\Big]\,.
\eea
The structure of the MST solutions is better understood by examining them in detail.
For example, Let us consider the MST, In solution for $\ell=2$.
We find
\bea
R_{\rm MST, in}^{\ell=2}(r)&=& c_2^{\eta_v^0}r^2+(c_1^{\eta_v^2}r+c_4^{\eta_v^2}r^4)\eta_v^2+
 c_2^{\eta_v^3}r^2\eta_v^3+( c_0^{\eta_v^4}+ c_3^{\eta_v^4}r^3+c_6^{\eta_v^4}r^6)\eta_v^4
 +(c_1^{\eta_v^5}r+ c_4^{\eta_v^5}r^4)\eta_v^5\nonumber\\
 &+&[ (c_2^{\eta_v^6}+ c_2^{\ln \eta_v^6}\ln (r))r^2 +c_5^{\eta_v^6} r^5 +c_8^{\eta_v^6}r^8]\eta_v^6+O(\eta_v^7)
\eea
where we summarize in the following Table the values of the various constants $c_n^{\eta_v^m}$.
\begin{table}  
\caption{\label{tab:table1}  List of the coefficients entering the solution $R_{\rm MST, in}^{\ell=2}$. The convenient notation $X=\frac{r_s}{r_b-r_s}$ and $LN=\ln\left((r_b-r_s)\eta_v^2\right)$ has been used. }
\begin{ruledtabular}
\begin{tabular}{l|l}
 $c_2^{\eta_v^0}$   & $ \frac{6}{(r_b-r_s)}$\\
 $c_1^{\eta_v^2}$   & $-\frac{6(r_b+r_s)}{(r_b-r_s)^2}$\\ 
 $c_4^{\eta_v^2}$   & $ -\frac{3\omega^2}{7(r_b-r_s)^2}$\\ 
 $c_2^{\eta_v^3}$   & $\omega\left(\frac{3 i 
   \left(r_b+r_s\right)}{\left(r_b-r_s\right)^2}-\frac{6 \text{LN} 
   r_s^{3/2}}{\left(r_b-r_s\right){}^{5/2}}\right)$\\ 
 $c_0^{\eta_v^4}$   & $ \frac{r_b^2+4r_br_s+r_s^2}{(r_b-r_s)^2}$\\ 
 $c_3^{\eta_v^4}$   & $ -\frac{(r_b+8r_s)\omega^2}{7(r_b-r_s)^2}$\\ 
 $c_6^{\eta_v^4}$   & $ \frac{\omega^4}{84(r_b-r_s)^2}$\\ 
 $c_1^{\eta_v^5}$   & $ \omega  \left(\frac{6 \text{LN} r_s^{3/2}
   \left(r_b+r_s\right)}{\left(r_b-r_s\right)^{5/2}}-\frac{3
   i
   \left(r_b+r_s\right)^2}{\left(r_b-r_s\right)^2}\right)=-(r_b+r_s)c_2^{\eta_v^3}$\\ 
 $c_4^{\eta_v^5}$   & $ \omega ^3 \left(\frac{3 \text{LN} r_s^{3/2}}{7
   \left(r_b-r_s\right)^{5/2}}-\frac{3 i
   \left(r_b+r_s\right)}{14 \left(r_b-r_s\right)^2}\right)=-\frac{\omega^2}{14}c_2^{\eta_v^3}$\\ 
 $c_2^{\eta_v^6}$   & $\omega ^2 \left(3 \text{LN}^2 X^3+\text{LN} \left(-6 i X^{5/2}-3 i X^{3/2}+\frac{18 X^2}{5}+\frac{9 X}{5}+\frac{16}{35}\right)-\pi ^2
   X^3+\frac{15 X^3}{2}-\frac{27 X^2}{35}-\frac{6 X}{35}+\frac{43}{735}\right)$\\ 
 $c_2^{\ln \eta_v^6}$   & $ -\frac{18 X^2}{5}-\frac{9 X}{5}-\frac{16}{35}$\\ 
 $c_5^{\eta_v^6}$   & $ \frac{\omega ^4 \left(2 r_b+7 r_s\right)}{84
   \left(r_b-r_s\right)^2}$\\ 
 $c_8^{\eta_v^6}$   & $ -\frac{\omega ^6}{5544 \left(r_b-r_s\right)^2}$\\
\end{tabular}
\end{ruledtabular}
\end{table}

\section{Expansions of $\Psi_{{\alpha}}$}\label{appc}
In this Appendix, we show some higher order terms in the double expansion in $q_1$ and $q_2$ of \eqref{z41nonexp}. Setting ${\alpha}=\pm$, the two independent solution of the wave equation in the AGT approach ca be expressed in the following form
\bea
\Psi_{{\alpha}}&=&z^{{\alpha} a+\frac{1}{2}}\Bigg\{1+\frac{k_0^2-p_0^2+a^2-\frac{1}{4}}{1+2{\alpha} a}q_1-\frac{c}{1-2{\alpha} a}q_2\nn\\
&+&\frac{8 a \left(2 a^3+4 a^2 {\alpha} +a-2 {\alpha} \right)+8
   k_0^2 \left(4 a (a+2 {\alpha} )-4 p_0^2+3\right)-8 p_0^2 (2
   a {\alpha} +1)^2+16 k_0^4+16 p_0^4-7}{64 (a {\alpha} +1) (2
   a {\alpha} +1)}q_1^2+\nn\\
&-&\frac{2 a {\alpha} -4 c^2-1}{2 (4 a-3 {\alpha} -1) (4 a-3
   {\alpha} +1)}q_2^2{+}\frac{c \left(4 k_0^2 (2 a {\alpha} +3)-4 p_0^2 (2 a {\alpha}
   +3)+(2 a {\alpha} -3) (2 a {\alpha} +1)^2\right)}{4 (2 a
   {\alpha} -1) (2 a {\alpha} +1)^2}q_1 q_2+\nn\\
   &+&\frac{q_1^3}{768(1{+}{\alpha} a)(1{+}2{\alpha} a)(3{+}2{\alpha} a)}\Big[(2 a {\alpha} {+}3)^2 \left(4 a \left(4 a \left(a^2{+}3 a {\alpha}
   {+}1\right){-}9 {\alpha} \right){-}17\right){+}16 k_0^4 \left(12 a
   (a{+}4 {\alpha} ){-}12 p_0^2{+}37\right)\nn\\
   &{+}&4 k_0^2 \left({-}24 p_0^2 (2
   a {\alpha} {+}3)^2{+}24 a (a (2 a (a{+}6 {\alpha} ){+}25){+}20 {\alpha}
   ){+}48 p_0^4{+}115\right){+}4 p_0^2 (p_0^2 (48 a (a{+}2 {\alpha}
   ){+}68)\nn\\
   &{-}&24 a (a (2 a (a{+}4 {\alpha} ){+}11){+}5 {\alpha} ){-}16
   p_0^4{-}7){+}64 k_0^6\Big]{+}\frac{q_1^2 q_2}{8(1{+}2{\alpha} a)^2(4a{-}3{+}{\alpha})(4a{+}3{+}{\alpha})}\Big[c (8 p_0^2 ({-}2 a \left(4 a^2 {\alpha} {+}6 a{-}5 {\alpha}
   \right)\nn\\
   &{+}&2 p_0^2 (2 a {\alpha} {+}5){+}11){+}(2 a {\alpha}
   {+}1)^2 \left(8 a^3 {\alpha} {-}4 a^2{-}18 a {\alpha} {-}3\right){+}8
   k_0^2 \left({-}4 p_0^2 (2 a {\alpha} {+}5){+}2 a (2 a (2 a {\alpha}
   {+}5){+}7 {\alpha} ){-}1\right)\nn\\
   &{+}&16 k_0^4 (2 a {\alpha} {+}5))\Big]{+}\frac{q_1q_2^2}{64({\alpha} a{-}1)(2{\alpha} a{-}1)^3(2{\alpha} a{+}1)^2}\Big[4 k_0^2 (4 a \left({\alpha}  \left(4 a^2 \left(2
   c^2{+}3\right){-}34 c^2{-}3\right){-}4 a \left(a^2{-}3
   c^2{+}1\right)\right)\nn\\
   &{+}&20 c^2{+}5){+}4 p_0^2 (4 a
   \left({\alpha}  \left({-}4 a^2 \left(2 c^2{+}3\right){+}34
   c^2{+}3\right){+}4 a \left(a^2{-}3 c^2{+}1\right)\right){-}5 \left(4
   c^2{+}1\right))\nn\\
   &{-}&\left(\left(1{-}4 a^2\right)^2 (2 a
   {\alpha} {-}5) \left(2 a {\alpha} {-}4 c^2{-}1\right)\right)\Big]{+}q_2^3\frac{c(5+4c^2-6{\alpha} a)}{12{\alpha}(2{\alpha} a-3)(2{\alpha} a-1)({\alpha} a-1)}\Bigg\}\,.
\eea
The result can be compared with the expression one finds in the MST approach, at least for the In solution. The dictionary between the two approaches is displayed in \eqref{dict1}, \eqref{dict2} and \eqref{dict3}.

\end{widetext}

\end{document}